\documentstyle[12pt,epsf]{article}
\newcommand{\nc}{\newcommand}
\nc{\renc}{\renewcommand}
\renc{\baselinestretch}{1.1}
\nc{\com}[1]{\ \\ \ {\bf \# {#1}}\\ \ }

\nc{\bort}[1]{}
\newlength{\overeqskip}
\newlength{\undereqskip}
\nc{\be}[1]{\begin{equation} \mbox{$\label{#1}$}}
\nc{\bea}[1]{\begin{eqnarray} \mbox{$\label{#1}$}}
\nc{\Section}[2]{\section{\sc #2}\label{#1}\seqnoll}
\nc{\Subsection}[2]{\subsection{\sc #2}\label{#1}}
\nc{\Bibitem}[1]{\bibitem{#1}}
\nc{\Label}[1]{\label{#1}}

\nc{\eea}{\vspace{\undereqskip}\end{eqnarray}}
\nc{\ee}{\vspace{\undereqskip}\end{equation}}
\nc{\bdm}{\begin{displaymath}}
\nc{\edm}{\end{displaymath}}
\nc{\dpsty}{\displaystyle}
\nc{\bc}{\begin{center}}
\nc{\ec}{\end{center}}
\nc{\ba}{\begin{array}}
\nc{\ea}{\end{array}}
\nc{\bab}{\begin{abstract}}
\nc{\eab}{\end{abstract}}
\nc{\btab}{\begin{tabular}}
\nc{\etab}{\end{tabular}}
\nc{\bit}{\begin{itemize}}
\nc{\eit}{\end{itemize}}
\nc{\ben}{\begin{enumerate}}
\nc{\een}{\end{enumerate}}
\nc{\bfig}{\begin{figure}}
\nc{\efig}{\end{figure}}
\nc{\seqnoll}{\setcounter{equation}{0}}
\renc{\theequation}{\thesection.\arabic{equation}}

\nc{\refc}[1]{\mbox{Ref.~\cite{#1}}}
\nc{\refs}[1]{\mbox{Refs.~\cite{#1}}}

\nc{\eqs}[2]{\mbox{Eqs.~(\ref{#1}) and (\ref{#2})}}
\nc{\eq}[1]{\mbox{Eq.~(\ref{#1})}}

\nc{\figs}[2]{\mbox{Figs.~\ref{#1} and \ref{#2}}}
\nc{\fig}[1]{\mbox{Fig.~\ref{#1}}}

\nc{\figcap}[1]{\begin{quote}\refstepcounter{figure}
        {\bf Figure \thefigure}: {\small\sl #1}\end{quote}}
\nc{\tabcap}[1]{\refstepcounter{table}
        {\bf Table \thetable}: {\small\sl #1}}

\nc{\tag}[1]{\label{#1} \marginpar{{\footnotesize #1}}}
\nc{\mtag}[1]{\label{#1} \mbox{\marginpar{{\footnotesize #1}}}}

\nc{\etal}{\mbox{\it et al. }}
\nc{\ie}{{\rm i.e. }}
\nc{\eg}{{\it e.g. }}

\nc{\arreq}{&\!\!\!=\!\!\!&}
\nc{\arrmi}{&\!\!\!!-\!\!\!&}
\nc{\arrpl}{&\!\!\!+\!\!\!&}
\nc{\arrap}{&\!\!\!\approx\!\!\!&}
\nc{\non}{\nonumber}
\nc{\nn}{\nonumber\\}
\nc{\align}{\!\!\!\!\!\!\!\!&&}

\nc{\mat}[4]{{\left(\ba{cc} #1 & #2 \\ #3 & #4 \ea\right)}}

\def\simleq{\; \raise0.3ex\hbox{$<$\kern-0.75em
      \raise-1.1ex\hbox{$\sim$}}\; }
\def\simgeq{\; \raise0.3ex\hbox{$>$\kern-0.75em
      \raise-1.1ex\hbox{$\sim$}}\; }
\nc{\DOT}{\hspace{-0.08in}{\bf .}\hspace{0.1in}}
\nc{\Laada}{\hbox {$\sqcap$ \kern -1em $\sqcup$}}
\nc\loota{{\scriptstyle\sqcap\kern-0.55em\hbox{$\scriptstyle\sqcup$}}}
\nc\Loota{{\sqcap\kern-0.65em\hbox{$\sqcup$}}}
\nc\laada{\Loota}
\nc{\qed}{\hskip 3em \hbox{\BOX} \vskip 2ex}

\nc{\real}{{\rm I \! R}}
\nc{\Z}{{\sf Z \!\!\! Z}}
\nc{\complex}{{\rm C\!\!\! {\sf I}\,\,}}
\def\diag{{\rm diag}}
\def\bigid{\leavevmode\hbox{\small1\kern-3.8pt\normalsize1}}
\def\id{\leavevmode\hbox{\small1\kern-3.3pt\normalsize1}}
\nc{\slask}{\!\!\!\!/}
\nc{\sslask}{\!\!\!/}
\nc{\bis}{{\prime\prime}}
\nc{\pa}{\partial}
\nc{\na}{\nabla}
\def\>{\rangle}
\def\<{\langle}
\nc{\goto}{\rightarrow}
\nc{\swap}{\leftrightarrow}

\nc{\EE}[1]{ \mbox{$\cdot10^{#1}$} }
\nc{\abs}[1]{\left|#1\right|}
\nc{\at}[2]{\left.#1\right|_{#2}}
\nc{\norm}[1]{\|#1\|}
\nc{\abscut}[2]{\abs{#1}_{\scriptscriptstyle#2}}
\nc{\vek}[1]{\hbox{\boldmath$#1$}}
\nc{\integral}[2]{\int\limits_{#1}^{#2}}
\nc{\inv}[1]{\frac{1}{#1}}
\nc{\dd}[2]{{{\partial #1}\over{\partial #2}}}
\nc{\ddd}[2]{{{{\partial}^2 #1}\over{\partial {#2}^2}}}
\nc{\dddd}[3]{{{{\partial}^2 #1}\over
        {\partial #2 \partial #3}}}
\nc{\dder}[2]{{{d #1}\over{d #2}}}
\nc{\ddder}[2]{{{d^2 #1}\over{d {#2}^2}}}
\nc{\dddder}[3]{{d^2 #1}\over
        {d #2 d #3}}
\nc{\dx}[1]{d\,^{#1}x}
\nc{\dy}[1]{d\,^{#1}y}
\nc{\dz}[1]{d\,^{#1}z}
\nc{\dl}[1]{\frac{d\,^{#1}l}{(2\pi)^{#1}}}
\nc{\dk}[1]{\frac{d\,^{#1}k}{(2\pi)^{#1}}}
\nc{\dq}[1]{\frac{d\,^{#1}q}{(2\pi)^{#1}}}
\nc{\dbar}{d\!\!\!\stackrel{\stackrel{\!-}{}}{}\!\!\!}
\nc{\cc}{\mbox{$c.c.$ }}
\nc{\hc}{\mbox{$h.c.$ }}
\nc{\cf}{cf.\ }
\nc{\erfc}{{\rm erfc}}
\nc{\Tr}{{\rm Tr\,}}
\nc{\tr}{{\rm tr\,}}
\nc{\pol}{{\rm pol}}
\nc{\sign}{{\rm sign}}
\nc{\bfT}{{\bf T }}

\nc{\cA}{{\cal A}}
\nc{\cB}{{\cal B}}
\nc{\cD}{{\cal D}}
\nc{\cE}{{\cal E}}
\nc{\cF}{{\cal F}}
\nc{\cG}{{\cal G}}
\nc{\cH}{{\cal H}}
\nc{\cL}{{\cal L}}
\nc{\cM}{{\cal M}}
\nc{\cO}{{\cal O}}
\nc{\cT}{{\cal T}}
\nc{\1}{\aa}
\nc{\2}{\"{a}}
\nc{\3}{\"{o}}
\nc{\4}{\AA}
\nc{\5}{\"{A}}
\nc{\6}{\"{O}}
\nc{\al}{\alpha}
\nc{\Del}{\Delta}
\nc{\e}{\epsilon}
\nc{\eps}{\epsilon}
\nc{\g}{\gamma}
\nc{\lam}{\lambda}
\nc{\om}{\omega}
\nc{\Om}{\Omega}
\nc{\ve}{\varepsilon}
\nc{\mn}{{\mu\nu}}
\nc{\ka}{\kappa}
\def\t{\tau}
\nc{\vp}{\varphi}
\nc{\pub}[4]{\Bibitem{#1}#2, {\sl ``#3''}, #4.}
\nc{\advp}[3]{{\it  Adv.\ in\ Phys.\ }{{\bf #1} {(#2)} {#3}}}
\nc{\annp}[3]{{\it  Ann.\ Phys.\ (N.Y.)\ }{{\bf #1} {(#2)} {#3}}}
\nc{\apl}[3]{{\it  Appl. Phys. Lett. }{{\bf #1} {(#2)} {#3}}}
\nc{\apj}[3]{{\it  Ap.\ J.\ }{{\bf #1} {(#2)} {#3}}}
\nc{\apjl}[3]{{\it  Ap.\ J.\ Lett.\ }{{\bf #1} {(#2)} {#3}}}
\nc{\app}[3]{{\it Astropart.\ Phys.\ }{{\bf #1} {(#2)} {#3}}}
\def\cjp#1#2#3{{\it  Can.\ J.\ Phys.\ }{{\bf #1} {(#2)} {#3}}}
\nc{\cmp}[3]{{\it  Comm.\ Math.\ Phys.\ }{{ \bf #1} {(#2)} {#3}}}
\nc{\cqg}[3]{{\it  Class.\ Quant.\ Grav.\ }{{\bf #1} {(#2)} {#3}}}
\nc{\epl}[3]{{\it  Europhys.\ Lett.\ }{{\bf #1} {(#2)} {#3}}}
\nc{\ijmp}[3]{{\it Int.\ J.\ Mod.\ Phys.\ }{{\bf #1} {(#2)} {#3}}}
\nc{\ijtp}[3]{{\it Int.\ J.\ Theor.\ Phys.\ }{{\bf #1} {(#2)} {#3}}}
\nc{\jmp}[3]{{\it  J.\ Math.\ Phys.\ }{{ \bf #1} {(#2)} {#3}}}
\nc{\jpa}[3]{{\it  J.\ Phys.\ A\ }{{\bf #1} {(#2)} {#3}}}
\nc{\jpc}[3]{{\it  J.\ Phys.\ C\ }{{\bf #1} {(#2)} {#3}}}
\nc{\jpg}[3]{{\it J.~Phys.~G:~Nucl.~Part.~Phys.~}{{\bf #1} {(#2)} {#3}}}
\nc{\jap}[3]{{\it J.\ Appl.\ Phys.\ }{{\bf #1} {(#2)} {#3}}}
\nc{\jpsj}[3]{{\it J.\ Phys.\ Soc.\ Japan\ }{{\bf #1} {(#2)} {#3}}}
\nc{\lmp}[3]{{\it Lett.\ Math.\ Phys.\ }{{\bf #1} {(#2)} {#3}}}
\nc{\lncim}[3]{{\it  Lett.\ Nuov.\ Cim.\ }{{\bf #1} {(#2)} {#3}}}
\nc{\mpl}[3]{{\it  Mod.\ Phys.\ Lett.\ }{{\bf #1} {(#2)} {#3}}}
\nc{\ncim}[3]{{\it  Nuov.\ Cim.\ }{{\bf #1} {(#2)} {#3}}}
\nc{\np}[3]{{\it  Nucl.\ Phys.\ }{{\bf #1} {(#2)} {#3}}}
\nc{\pr}[3]{{\it Phys.\ Rev.\ }{{\bf #1} {(#2)} {#3}}}
\nc{\pra}[3]{{\it  Phys.\ Rev.\ }{{\bf A#1} {(#2)} {#3}}}
\nc{\prb}[3]{{\it  Phys.\ Rev.\ }{{\bf B#1} {(#2)} {#3}}}
\nc{\prc}[3]{{\it  Phys.\ Rev.\ }{{\bf C#1} {(#2)} {#3}}}
\nc{\prd}[3]{{\it  Phys.\ Rev.\ }{{\bf D#1} {(#2)} {#3}}}
\nc{\prl}[3]{{\it Phys.\ Rev.\ Lett.\ }{{\bf #1} {(#2)} {#3}}}
\nc{\pl}[3]{{\it  Phys.\ Lett.\ }{{\bf #1} {(#2)} {#3}}}
\nc{\prep}[3]{{\it Phys\. Rep.\ }{{\bf #1} {(#2)} {#3}}}
\nc{\prsl}[3]{{\it Proc.\ R.\ Soc.\ London\ }{{\bf #1} {(#2)} {#3}}}
\nc{\ptp}[3]{{\it  Prog.\ Theor.\ Phys.\ }{{\bf #1} {(#2)} {#3}}}
\nc{\ptps}[3]{{\it  Prog\ Theor.\ Phys.\ suppl.\ }{{\bf #1} {(#2)} {#3}}}
\nc{\physa}[3]{{\it  Physica\ A\ }{{\bf #1} {(#2)} {#3}}}
\nc{\physb}[3]{{\it  Physica\ B\ }{{\bf #1} {(#2)} {#3}}}
\nc{\phys}[3]{{\it Physica\ }{{\bf #1} {(#2)} {#3}}}
\nc{\rmp}[3]{{\it  Rev.\ Mod.\ Phys.\ }{{\bf #1} {(#2)} {#3}}}
\nc{\rpp}[3]{{\it Rep.\ Prog.\ Phys.\ }{{\bf #1} {(#2)} {#3}}}
\nc{\sjnp}[3]{{\it Sov.\ J.\ Nucl.\ Phys.\ }{{\bf #1} {(#2)} {#3}}}
\nc{\spjetp}[3]{{\it Sov.\ Phys.\ JETP\ }{{\bf #1} {(#2)} {#3}}}
\nc{\yf}[3]{{\it Yad.\ Fiz.\ }{{\bf #1} {(#2)} {#3}}}
\nc{\zetp}[3]{{\it Zh.\ Eksp.\ Teor.\ Fiz.\ }{{\bf #1} {(#2)} {#3}}}
\nc{\zp}[3]{{\it Z.\ Phys.\ }{{\bf #1} {(#2)} {#3}}}
\nc{\zpc}[3]{{\it Z.\ Phys.\ C\ }{{\bf #1} {(#2)} {#3}}}
\nc{\ibid}[3]{{\sl ibid.\ }{{\bf #1} {#2} {#3}}}
\newcommand{\minus}{\!-\!}
\newcommand{\plus}{\!+\!}
\nc{\Leff}{{\cal L}^{\beta,\mu}_{\rm eff}}
\nc{\dLeff}{\Delta{\cal L}^{\beta,\mu}_{\rm eff}}
\def\vPi{\vek{\Pi}}
\def\vp{\vek{p}}
\def\vx{\vek{x}}
\def\vy{\vek{y}}
\def\vg{\vek{\g}}
\def\Psibar{\overline{\Psi}}

\def\LLL{lowest Landau level}
\def\mme{\cM^2_e}
\begin{document}
\baselineskip 16pt
%
\thispagestyle{empty}
\begin{flushright} 
  CERN-TH/96-207\\
  hep-th/9608271\\
\end{flushright}
\begin{center}
{\Huge\bf   Hard Thermal Loops \\
in a Magnetic Field\\[4mm]
and the Chiral Anomaly}
\normalsize
\end{center}
%
\bc
{\bf\large Per Elmfors}
 \\[2mm]
{\sl CERN, TH-Division,
CH-1211 Geneva 23, Switzerland \\ 
elmfors@cern.ch}
\ec

%
\vfill
\bc
{\bf Abstract} \\
\ec
{\small
\begin{quotation}
\noindent
The   fermionic dispersion relation in  the   presence of a background
magnetic field and a high temperature QED plasma is calculated exactly
in the external field,  using the Hard  Thermal Loop  effective action.
As the field strength increases there is a  smooth transition from the
weak-field ($qB\ll q^2T^2$) thermal dispersion relations to the vacuum
Landau levels when   the background field  is  much stronger  than any
thermal  effects  ($qB\gg q^2T^2$).  The   self-energy at finite field
strength   acquires  an imaginary part.    The  spectral width becomes
important   for  critical  field   strengths   ($qB   \sim   q^2T^2$),
necessitating the use of the full  spectral function. It is shown that
the spectral function satisfies the usual condition of normalization and
causality.  Using the exact spectral  function I  also show that  the
production of  chirality in an  external electromagnetic field at high
temperature is  unaffected by the  presence  of the thermal masses  of the
fermions. 
\end{quotation}}
\vfill
\begin{flushleft} 
  CERN-TH/96-207\\
  July 1996\\
\end{flushleft}

\newpage
\normalsize
\setcounter{page}{1}
\baselineskip 18pt
%
\Section{s:intro}{Introduction}
During   the  last  few  years  an  increasing   understanding of high
temperature gauge theories   has been achieved through  the consistent
resummation of leading  diagrams  called   Hard Thermal Loops    (HTL)
\cite{BraatenP90,FrenkelTW90}.  The HTL resummed  theory has been used
to improve  IR divergences in a  number of  processes, the most famous
one  being thermal gluon   scattering  and the  related gluon  damping
problem  \cite{BraatenP90b}.     The HTL effective    action  has been
constructed in several different ways, and it yields a gauge-invariant
extension of the electric screening mass for non-static fields.  Since
strong background fields are likely to  have been present in the early
Universe it is of interest to look for  solutions to the effective HTL
equations of  motion in such  backgrounds.  Most  of the literature so
far has   been concentrated on the small   fluctuations  around the trivial
background with zero  field.  It has been  known for a long time  that
static magnetic fields are   not screened by  the HTL  resummation and
thus a constant  magnetic  field is also   a solution to the  resummed
effective equation of  motion.  In this paper  I explore the fermionic
excitations around such a constant field at very high temperature. One
interesting point is that this can be done exactly, i.e. to all orders
in the external  field.  At the end we  recover two well-known limits.
On the one   hand there is 
the zero-field     limit where the  thermal  dispersion
relations  are well known   \cite{Klimov82,Weldon8289}.  On the  other
hand, when  the field  is  very  strong  compared with  the temperature,
thermal corrections become less important and in that limit we recover
the zero-temperature  Landau levels.  For intermediate field strengths
the self-energy   acquires  a non-negligible  imaginary  part   due to
synchrotron radiation and scattering   with the heat bath,  even above
the light-cone.  It    is,  therefore, necessary to   study   the full
spectral function and not only an on-shell relation for the real part.
It  turns out that in the  lowest Landau level   the spectral width is
rather large, when the field strength gets  comparable with  the thermal
mass squared ($qB\sim \cM^2$), and the quasi-particle picture is not
reliable.  

Since there is a mass gap in the thermal fermionic spectrum even for a
chirally invariant theory, it is not immediately  obvious how the chiral
anomaly mechanism works. The  standard  level crossing picture is  not
applicable  since no  levels ever  cross  the Dirac surface.  I  show,
using the full spectral function in a background of electric and magnetic
fields, how   the  spectral   weight  associated with    particles  and
antiparticles can move  continuously between the positive and negative
energy  solution without crossing the Dirac  surface, and through this
mechanism satisfy the anomaly equation. 
 
In section \ref{s:dr}, I describe  the method of diagonalizing the
HTL effective action  in a background  magnetic field, and compare the
exact  result with an   approximate   weak field  formula  in  section
\ref{s:comp}.   The   spectral function    is  calculated   in section
\ref{s:spfkn}.   The main issue  of the paper,   namely the anomaly 
mechanism at
finite temperature,  is treated  in section
\ref{s:chiral} for the  case  of a free  field in 1+1
dimension  and in  the HTL  approximation in 3+1  dimension.  
Some properties  such as  normalization and  causality
are discussed in an appendix. 

\Section{s:dr}{Dispersion relations from Hard Thermal Loops}
The HTL effective action for 
QED can be written as~\cite{Braaten93}
\bea{HTLEA}
    \cL_{\rm HTL}=\align-\inv{4}F^2+\frac{3}{4}\cM^2_\gamma 
    F_{\mu\alpha}\left<\frac{u^\alpha u^\beta}
      {(\pa u)^2}\right> F_\beta^{~~\mu}\nn
      \align +\,\,\Psibar(\Pi\slask-m)\Psi
    -\cM_e^2\Psibar\gamma_\mu\left<\frac{u^\mu}{u\cdot\Pi}
      \right>\Psi~~,
\eea
where $\Pi_\mu=i\pa_\mu-qA_\mu$ and 
the average $\left<\cdot\right>$ is defined by
\be{ave}
    \left< f(u_0,\vek{u})\right>=\int \frac{d\Omega}{4\pi}
      f(u_0,\vek{u})~~,
\ee
where $u_0=1$ and  $\vek{u}$ is a   spatial unit vector.  The  thermal
mass of the  photon $\cM^2_\gamma$ is  given by $q^2T^2/9$ and for the
electron  we have  $\cM_e^2=q^2T^2/8$.  The  equation   of motion  for
$\Psi$    that follows      from    \eq{HTLEA}    is      
\be{eqom}
    \left[\Pi\slask-m-\cM_e^2                                   
     \gamma_\mu
     \left<\frac{u^\mu}{u\cdot\Pi}\right>\right]\Psi=0~~.   
\ee    
Equation
(\ref{eqom})  is a  non-local  and  non-linear differential  equation,
which  is, in   general, very  difficult  to solve.   What makes  this
equation much less tractable than  the  thermal Dirac equation, in  the
absence of an external electromagnetic  field, is that the average over
$\vek{u}$     is        difficult       to          perform explicitly since
$[\Pi_\mu,\Pi_\nu]=-ieF_{\mu\nu}\neq  0$,\  \ie  not all components of
$\Pi_\mu$   can  be diagonalized  simultaneously.    We  shall in this
section only deal with an external magnetic field and fix  it to be in
the $z$-direction. The solutions to \eq{eqom} in vacuum ($\cM_e=0$) are
given by the standard Landau levels.  Since  the spatial symmetries of
the system are unchanged by the thermal heat bath, we expect the
eigenfunctions to   have    the   same  spatial   form    as  at  zero
temperature. In fact,  after performing the  $u$-integral in \eq{eqom}
the  result can  only be a  function  of the invariants $\Pi_\perp^2$,
$p_0^2$ and $p_z^2$, and the $\gamma$-structure has 
to be proportional to $\gamma\Pi_\perp$, $\gamma_0p_0$ and $\gamma_zp_z$.%
\footnote{
We use the notation $a\cdot b_\perp=a_xb_x+a_yb_y$ for any two 
four-vectors $a$ and $b$. In our convention three-vectors such as
$\vp=(p_x,p_y,p_z)$ and $\vg=(\g_x,\g_y,\g_z)$ 
are the contravariant components of the corresponding
four-vector and thus have  Lorentz indices $i=1,2,3$ upstairs,
i.e. $p_x=p^1$ etc. We use 
the Minkowski metric ${\rm diag}(+,-,-,-)$ so that $p_i=-p^i$ and
$\g_i=-\g^i$ for $i=1,2,3$.}

We shall therefore compute the matrix elements 
\be{matel}
    \<\Phi_{\kappa'}|\left<\frac{u^\mu}{u\cdot\Pi}\right>
    |\Phi_\kappa\>
\ee
between the vacuum eigenstates.
To be specific we use the gauge $A_\mu=(0,0,-Bx,0)$. Then the 
eigenstates are given by
\bea{state}
    \<x|\Phi_{\kappa}\>&=&\exp[i(- p_0 t \plus  p_{y}y \plus  p_{z}z ) ]
    I_{n;p_y}(x) ~~,\\[2mm]
    I_{n;p_y}(x)&=& \left(\frac{|qB|}{\pi} \right)^{1/4} \exp \left[
    - \inv{2} |qB| \left( x \minus \frac{p_y}{qB} \right)^{2}\right]\nn
    &&\times\inv{\sqrt{n!}} H_n \left[\sqrt{2|qB|} \left(x-\frac{p_y}
    {qB} \right) \right] ~~,
\eea
where $\kappa=\{p_0,n,p_y,p_z\}$ and $H_n[x]$ are Hermite polynomials defined
by
\be{herm}
    H_n[x]=(-1)^n e^{\frac{x^2}{2}}\frac{d^n}{dx^n}e^{-\frac{x^2}{2}}~~.
\ee
These states form a complete set of functions in four dimensions when the
energy is off shell. 
In the chiral representation and with $qB>0$, 
suitable spinors can be formed from 
$\Phi_\kappa$ as 
$\Psi_\kappa={\rm diag}[\Phi_\kappa,\Phi_{\kappa-1},\Phi_\kappa,%
\Phi_{\kappa-1}]\chi_\kappa$ where $\chi_\kappa$ is a
space-time-independent spinor, which can be determined from the
Dirac equation.
The vacuum Dirac operator in \eq{eqom} gives by construction
an eigenvalue when acting on $\Psi_\kappa$, but it is more
difficult to determine the action of the thermal part since $\Phi_\kappa$
cannot be an eigenfunction to $u\cdot\Pi$ for all $u$. One way of calculating
the matrix element in \eq{matel} is to find a basis such that 
$v\cdot\Pi|v_p\>=v\cdot p|v_p\>$ and insert a unit operator
$\int [d^4p]|v_p\>\<v_p|$ into \eq{matel}. 
(We use the notation $\int [d^np]=\int d^np/(2\pi)^n$.) The unit operator
is, of course, independent of $v$, so
in particular we can choose $v=u$ and change the order of integrations
between $p$ and $u$.
In the gauge we use, an eigenvector to $v\cdot \Pi$ is given by
\be{vvector}
    \< x|v_p\>=\exp\left[-ip_0t+ip_zz+ip_yy+
    i\left(p_xx+\frac{qB v_y}{v_x}\frac{x^2}{2}\right)\right]~~.
\ee
After computing the matrix elements in \eq{matel} we find indeed
that they are diagonal in $\kappa$ for $u_0$ and $u_z$, and 
have a mixing with the first subdiagonals for $u_x$ and $u_y$.
We define $\<u_{0,z,\pm}\>$ by
\bea{u}
     \<\Phi_{\kappa'}|\left<\frac{u_{0,z}}{u\cdot\Pi}\right>|\Phi_\kappa\>
     &=& (2\pi)^3\delta_{\kappa',\kappa}\<u_{0,z}\>_\kappa~~,\\[2mm]
     \<\Phi_{\kappa'}|\left<\frac{u_x\pm iu_y}{u\cdot\Pi}\right>|\Phi_\kappa\>
     &=& (2\pi)^3\delta_{\kappa',\kappa\mp1}\<u_\pm\>_\kappa~~,
\eea
and $\kappa\mp1=\{p_0,n\mp1,p_y,p_z\}$. These are exactly the components
that occur naturally when we include the $\gamma$-matrices
in the chiral representation.
The explicit 
calculation of $\<u_{0,z,\pm}\>$ is a bit lengthy but straightforward
and is done by performing the integrals over $x$, $x'$, $p$ and $u$ in
\be{xxpu}
    \<\Phi_{\kappa'}|\left<\frac{u_\mu}{u\cdot\Pi}\right>|\Phi_\kappa\>
    =\int\!dx\,dx'\,dp\,\frac{d\Omega}{4\pi}
    \<\Phi_{\kappa'}|x'\>\<x'|u_p\>\frac{u_\mu}{u\cdot p}
    \<u_p|x\>\<x|\Phi_\kappa\>~~.
\ee
The result reads
\bea{u0}
    \<u_0\>_\kappa\arreq\inv{n!\sqrt{2\pi}}\int_{-\infty}^\infty
        ds\,H_n^2(s) e^{-s^2/2}
        \nn\align\times
        \left\{\frac{p_z}{2p^2}\ln\frac{p_0+p_z}{p_0-p_z}
          +\frac{p_0s\sqrt{2qB}}{2p^2\sqrt{p_0^2-p^2}}
          \arctan\frac{s\sqrt{2qB}}{2\sqrt{p_0^2-p^2}}\right\}~~,
\eea
\bea{uz}
    \<u_z\>_\kappa\arreq\inv{n!\sqrt{2\pi}}\int_{-\infty}^\infty
        ds\,H_n^2(s) e^{-s^2/2}
        \nn\align\times
        \left\{-\frac{p_z}{p^2}+
          \frac{p_0(2p_z^2-qBs^2)}{4p^4}\ln\frac{p_0+p_z}{p_0-p_z}
          \right.\nn\align\left.
          +\frac{p_z(2p_0^2-p^2)}{2p^4}
          \frac{s\sqrt{2qB}}{\sqrt{p_0^2-p^2}}
          \arctan\frac{s\sqrt{2qB}}{2\sqrt{p_0^2-p^2}}\right\}~~,
\eea
\bea{up}
    \<u_+\>_\kappa\arreq\frac{i}{\sqrt{2\pi n!(n-1)!}}\int_{-\infty}^\infty
        ds\,H_n(s)H_{n-1}(s) e^{-s^2/2}
        \nn\align\times
        \left\{\frac{s\sqrt{2qB}}{2p^2}
          -\frac{p_0sp_z\sqrt{2qB}}{2p^4}\ln\frac{p_0+p_z}{p_0-p_z}
          \right.\nn\align\left.
          +\frac{2p_z^2(p_0^2-p^2)-p_0^2 qB s^2}{2p^4\sqrt{p_0^2-p^2}}
          \arctan\frac{s\sqrt{2qB}}{2\sqrt{p_0^2-p^2}}\right\}~~,\\
    \<u_-\>_\kappa&=&-\<u_+\>_{\kappa+1}~~,
\label{um}
\eea
where $p^2=p_z^2+qBs^2/2$.
With these definitions the Dirac equation effectively reduces
to a $4\times4$ matrix in the spinor indices, since the other
quantum numbers have been diagonalized. 
In the massless limit ($m=0$) the left- and right-handed parts factorize
and the Dirac equation takes the form
\be{RLDirac}
    \left[\Pi\slask-\cM_e^2 \gamma_\mu
      \left<\frac{u^\mu}{u\cdot\Pi}\right>\right]\chi\equiv
    \mat{0}{D_L(\kappa)}{D_R(\kappa)}{0}
    \left(\ba{c} \chi_R(\kappa) \\ \chi_L(\kappa)\ea\right)=0~~,
\ee
where
\bea{DRDL}
    D_R(\kappa)&=&\left(\ba{cc} 
      -p_0+p_z+\cM_e^2(\<u_0\>_\kappa-\<u_z\>_\kappa) &
      i\sqrt{2qBn}-\cM_e^2\<u_-\>_{\kappa-1} \\
      -i\sqrt{2qBn}-\cM_e^2\<u_+\>_{\kappa} &
      -p_0-p_z+\cM_e^2(\<u_0\>_{\kappa-1}+\<u_z\>_{\kappa-1})
      \ea \right)~~,\nn
    D_L(\kappa)&=&\left(\ba{cc} 
      -p_0-p_z+\cM_e^2(\<u_0\>_\kappa+\<u_z\>_\kappa) &
      -i\sqrt{2qBn}+\cM_e^2\<u_-\>_{\kappa-1} \\
      i\sqrt{2qBn}+\cM_e^2\<u_+\>_{\kappa} &
      -p_0+p_z+\cM_e^2(\<u_0\>_{\kappa-1}-\<u_z\>_{\kappa-1})
      \ea \right)~~,\nn
\eea
In the lowest Landau level ($n=0$) \eq{RLDirac} reduces to a
$2\times2$ matrix, since only one orientation of 
the magnetic moment is possible.
It is easy to take the determinant of \eq{DRDL} to find the
dispersion relations, which for the right-handed component are
\be{exdr}
\ba{rrl}
    n\geq 1:~&\Bigl(p_0-p_z-\cM_e^2
    (\<u_0\>_\kappa-\<u_z\>_\kappa)\Bigr)&\\[2mm]
    &\times \Bigl(p_0+p_z-\cM_e^2
    (\<u_0\>_{\kappa-1}+\<u_z\>_{\kappa-1})\Bigr)&\\[2mm]
    &-\left(\sqrt{2qBn}-i\cM_e^2\<u_+\>_\kappa\right)^2&=~0~~,\\[2mm]
    n=0:~& p_0-p_z-\cM_e^2
    (\<u_0\>_\kappa-\<u_z\>_\kappa)&=~0~~.
\ea
\ee
\bfig[t]
\setlength{\unitlength}{1mm}
\begin{picture}(80,90)(0,0)
\includegraphics{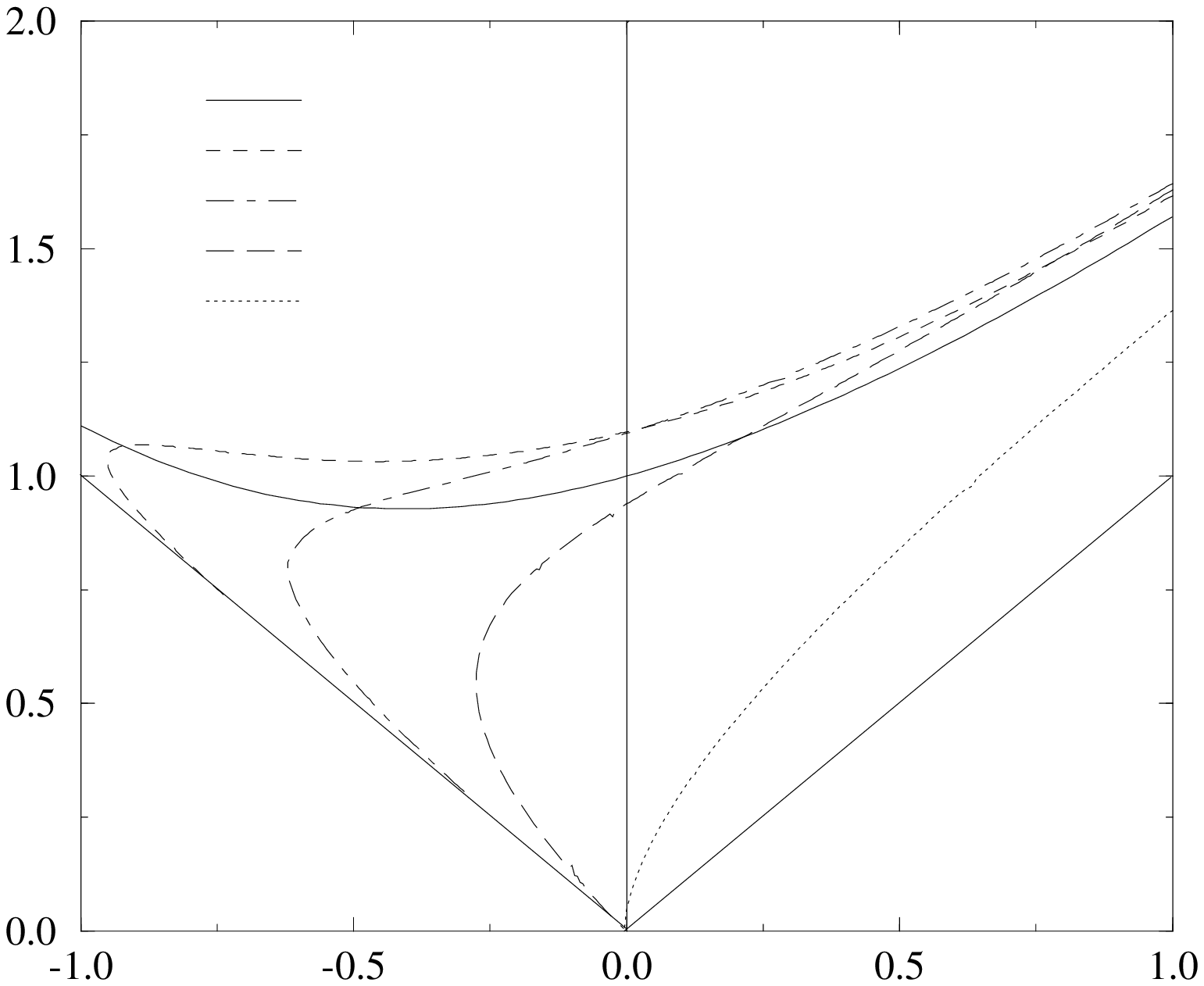}
   \put(20,0){}
   \put(65,-2){$p_z$}
   \put(17,50){$E$}
   \put(52,76){\small $qB=0.0$}
   \put(52,71.5){\small $qB=0.5$}
   \put(52,67){\small $qB=1.0$}
   \put(52,62.5){\small $qB=2.0$}
   \put(52,58){\small $qB=10.0$}
\end{picture}
   \figcap{Dispersion relations for the right-handed branch in 
     the lowest Landau level ($n=0$), neglecting the
     imaginary part. As the $B$-field increases
     thermal effects become less important and
     the dispersion relation approaches the light cone, which is 
     indicated by solid lines.
     All dimensionful parameters are given in units of the thermal
     mass $\cM_e$.
\label{f:n0cont}}
\efig
These relations are only meaningful for stable propagating quasi-particles
with well-defined relations between momentum and energy, \ie when the
imaginary parts are negligible.
In general there are imaginary parts in the functions
$\<u_{0,z,\pm}\>_\kappa$, which are discussed in
Section~\ref{s:spfkn}. 
It is anyway useful to first solve \eq{exdr}, ignoring for the moment
the imaginary part, since the zeros of the real part indicate where the
spectral functions are peaked, at least when the imaginary part
is small enough. This can conveniently be done numerically
as all the integrals in Eqs.~(\ref{u0}) to (\ref{um}) are well convergent.
The dispersion
relations for  several field strengths in the lowest Landau level are
shown in \fig{f:n0cont}. 

In the  lowest  Landau level     a  right-handed particle    (positive
chirality, $\chi=R=+1$) with $q$ and $B$ positive,  can only propagate in
the positive  $z$-direction  since the magnetic  moment,  and thus the
spin, has  to  point  along the   field.  Positive chirality   implies
positive helicity for particles and thus positive $p_z$. The left-hand
side of \fig{f:n0cont} has to violate one of these sign arguments. The
solution is  that  hole excitations  have opposite  chirality--helicity
relation   and  can thus propagate   for  negative $p_z$.  
As the  field increases the hole branch
develops a new sub-branch and disappears continuously for large enough
fields. The new branch must  not be taken  too seriously, since it only
appears when the  imaginary part is non-negligible  and then only  the
full spectral function  is meaningful.  The particle branch approaches
the light cone, i.e. the vacuum  dispersion relation, in a smooth  way
as  the field strength increases.  This  is physically very reasonable
since,  for very strong  field  strengths,  the thermal effects   should
disappear.  Once again it should be emphasized that the above analysis
is based only on  the  real part of  the  self-energy and it can  only
serve as a guiding line to describe what kind of  modes propagate in the
plasma.  For a complete description, which is necessary for $qB\simeq
\cM_e^2$, where the  imaginary part is comparable with  the real part,  the
full   spectral function    has   to    be    used, as    we  do    in
section~\ref{s:spfkn}. 

The HTL effective action is derived under the condition that the 
temperature is much larger then the momentum. Here, the magnetic
field enters only through the covariant momentum and should
thus satisfy the condition $\Pi^2\sim qB\ll T^2$. 
On the other hand, already when $qB\gg\cM_e^2\sim q^2T^2$
(which is the only scale where $T$ enters in the HTL approximation)
the thermal corrections start to get small compared with the tree-level
part. Thus, for small coupling the HTL corrections become small before
they are invalid.
\Section{s:comp}{Comparison with an approximate formula}
In a direct one-loop calculation of the fermionic self-energy
in a magnetic field \cite{ElmforsPS95} one is naturally
led to an approximation where the full $B$-dependence is kept only where
it is added linearly to the momentum squared. In other places it enters
only to $\cO(B^2)$ (see \cite{ElmforsPS95} for details). Since the
result from this approximation is surprisingly simple, it is worth
commenting on its relation with the exact solution. From 
\cite{ElmforsPS95} we find the Dirac equation
\bea{Deq}
        &&[\Pi\slask-m-\hat{\Sigma}(p_0,p_z,\vPi_\perp)]\Psi=\nn[2mm]
        &&\left[s(p_0,\vPi^2)\gamma_0p_0 -
        r(p_0,\vPi^2)\gamma_zp_z-
        r(p_0,(\vPi\cdot\vg)^2)\Pi\slask_\perp
        -m\right]\Psi=0~~,\quad
\eea
where $\vPi^2=\Pi_\perp^2+p_z^2$. The functions 
$s(p_0,\vPi^2)$ and $r(p_0,\vPi^2)$
are derived from the HTL effective action without background field
and they are given by:
\bea{s}
        p_0s(p_0,\vPi^2) &=& p_0-\cM_e^2
        \at{\left\<\frac{u_0}{u\cdot p}\right\>}{p\goto\Pi}
        =p_0-\frac{\cM_e^2}{2\abs{\vPi}}
        \ln\abs{\frac{p_0+\abs{\vPi}}{p_0-\abs{\vPi}}} \ ,
        \\[2mm]
        p_z r(p_0,\vPi^2) &=&p_z-\cM_e^2
        \at{\left\<\frac{u_z}{u\cdot p}\right\>}{p\goto\Pi}
        =p_z+\frac{p_z\cM_e^2}{\vPi^2}
        \left(1-\frac{p_0}{2\abs{\vPi}}
        \ln\abs{\frac{p_0+\abs{\vPi}}{p_0-\abs{\vPi}}}\right)\ .\nn
\label{r}
\eea
It is {\em almost} possible to guess the expression
in \eq{Deq} from the standard expression for the HTL Dirac 
equation~\cite{Klimov82,Weldon8289}.
The usual momentum $p_\mu$ should be replaced with the gauge-invariant
momentum $\Pi_\mu$, but there is an ambiguity in replacing $p^2$ by $\Pi^2$
or by $\Pi\slask\,\,\Pi\slask\,\,$. The correct way follows from the 
calculations in \cite{ElmforsPS95}.

The difference between \eq{Deq} and
the exact formula is related to the order of doing
the average over $u$ and replacing $p$ by $\Pi$. Comparing the exact expression
for $\<u_0\>$ with \eq{s}, before doing the $u$-integration, we would like
to specify under which circumstances the approximate equality
\bea{u0equ0}
    &&\inv{n!\sqrt{2\pi}}\int_{-\infty}^\infty
    ds\,H_n^2(s) e^{-s^2/2}
    \frac{d\Omega}{4\pi}
    \frac{u_0}{u_0p_0-u_zp_z-u_\perp
      \sqrt{\frac{qB}{2}}s}\nn
    &&\simeq
    \int\frac{d\Omega}{4\pi}\at{\frac{u_0}{u_0p_0-u_zp_z-
      u\cdot p_\perp}}{p_\perp\goto\Pi_\perp}
\eea
is valid. The first term comes from the exact expression and the second
from the approximate formula \eq{s}.
In a standard coordinate system with $\vek{u}=(\sin\theta\sin\phi, 
\sin\theta\cos\phi,\cos\theta)$, so that $u_\perp=\sin\theta$,
 we see that the difference lies in the
integral over the azimuthal angle $\phi$. In the exact formula the
integral over $\phi$ is replaced by a more complicated integral over
$s$, involving the exact external states, i.e. the Landau levels.
The two expressions in \eq{u0equ0} do not coincide, except in some 
particular limits.
Expanding both sides of \eq{u0equ0} formally in powers of $qB$ and 
$\Pi_\perp^2$, we are led to comparing the integrals
\bea{upperpcomp}
    \inv{n!\sqrt{2\pi}}\int_{-\infty}^\infty
    ds\,H_n^2(s) e^{-s^2/2}
    \left(u_\perp\sqrt{\frac{qB}{2}}s\right)^{2k}&=&
    (u_\perp^2qB)^k\frac{(2k)!}{2^{2k}}\sum_{l=0}^{{\rm min}(n,k)}
    \frac{2^ln!}{(n-l)!(k-l)!(l!)^2}~~,\nn
    \int_0^{2\pi}\frac{d\phi}{2\pi}
    \at{\left(\vek{u}_\perp\cdot p_\perp \right)^{2k}}{p_\perp\goto\Pi_\perp}
    &=&(u_\perp^2qB)^k \frac{(2k)!}{2^{2k}(k!)^2} (2n+1)^{k}~~,
\eea
where we used the fact 
that $\Pi_\perp^2=qB(2n+1)$ when acting on a Landau level.
First we notice that for $k=1$ the two integrals coincide.
Then, we find that the leading terms in the limit of large $n$, for fixed $k$,
also coincide. We can thus expect that the approximative formula
\eq{Deq} is useful both for weak fields and for very high Landau levels. 
It should, however, be noticed that the expansion converges badly close to the
light cone, and that it eventually breaks down for hole excitation of
high momentum \cite{ElmforsPS95}. 

In many applications it is only the dispersion relation for small momenta
that is important.
Using \eq{Deq}  in the limit $m=0$, 
we can easily obtain an approximate formula for the  dispersion relation
 in the lowest Landau level around $p_z=0$ 
in the presence of a weak magnetic field and $\chi=1$
\be{EthB}
    E(p_z)\simeq \cM_e\left(1+\frac{qB}{6\cM_e^2}\right)
    +\frac{p_z}{3}\left(1-\frac{7qB}{15\cM^2_e}\right)~~.
\ee
For the right-handed branch
of the \LLL{} in a weak magnetic field ($qB=0.2\,\cM_e^2$), the dispersion
relation following from \eq{Deq} is
shown in \fig{f:comp}, where it is also compared with the exact
solution of \eq{exdr}.
\bfig[t]
\setlength{\unitlength}{1mm}
\begin{picture}(85,90)(0,0)
\includegraphics{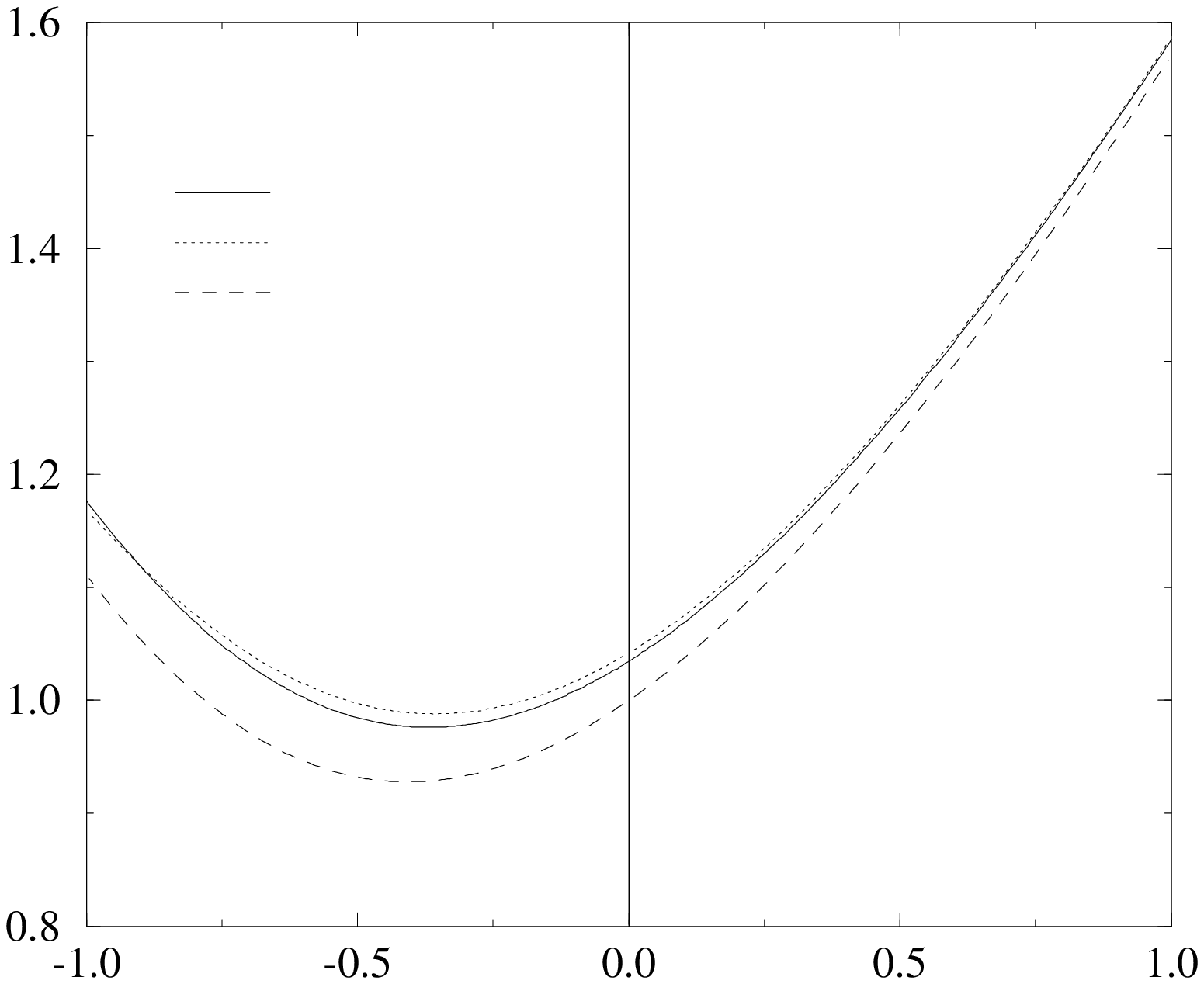}
   \put(0,0){}
   \put(90,0){$p_z$}
   \put(17,50){$E$}
   \put(50,68){\small Exact HTL}
   \put(50,64){\small Weak field}
   \put(50,60){\small $qB=0.0$}
\end{picture}
   \figcap{Comparison of the dispersion relation from the HTL
     effective action and the weak field approximation
     in the lowest Landau level for $qB=0.2\cM_e^2$.
     All dimensionful parameters are given in units of the thermal
     mass $\cM_e$.
\label{f:comp}}
\efig
%
\Section{s:spfkn}{Spectral function}
The dispersion relation was solved in section~\ref{s:dr},
ignoring the imaginary parts of the self-energy. This is only
a good approximation for small magnetic fields, where the imaginary
parts are small. Since we have the exact expression for the self-energy
it is not too difficult  to study 
the complete spectral function and to see how good the
quasi-particle picture is. 
The spectral function can be defined from
the representation of the retarded and advanced propagator as
\be{SeqA}
    S(E\pm i\eps,\vp)=\int_{-\infty}^\infty dE'
    \frac{\cA(E',\vp)}{E-E'\pm\,i\epsilon}~~.
\ee
In the real-time formalism of thermal field theory the spectral representation
goes through in much the same way as at zero temperature, the only 
essential
difference being doubling the degrees of freedom
(for a recent review see \cite{Henning95}). The full thermal propagator
takes the form \cite{Henning95}
\be{RTS}
   S^{(ab)}(E,\vp)=\int_{-\infty}^\infty dE'\cA(E',\vp)\sigma_z
   \cB^{-1}(E')
   \left(\ba{cc} \inv{E-E'+i\epsilon} & 0 \\
                 0 & \inv{E-E'-i\epsilon} \ea\right)
    \cB(E')~~,
\ee
where $\sigma_z$ is a Pauli spin matrix and $\cB(E)$ can be chosen to be
\be{B}
    \cB(E)=\left(\ba{cc} 
      (e^{-\beta(E-\mu)}+1)^{-1} & (e^{\beta(E-\mu)}+1)^{-1}\\
      1 & 1 \ea\right)~~.
\ee
For a free Dirac fermion we have
\be{Afree}
   \cA(E,\vp)=(\g_0E-\vg\vp+m)\sign(E)
   \delta(E^2-\vp^2-m^2)~~,
\ee
but in general $\cA(E,\vp)$ can have both a $\delta$-function
part for the quasi-particles and a continuous part.
The HTL fermion propagator without any external $B$-field 
is given by
\be{SHTL}
    S(E,\vp)=\frac{s(E,\vp)\g_0E-r(E,\vp)\vg\vp+m}
    {s(E,\vp)^2E^2-r(E,\vp)^2\vp^2-m^2}~~,
\ee
where the functions $s(E,\vp)$ and $r(E,\vp)$ are defined in
\eqs{s}{r}. For $E>\abs{\vp}$ the only imaginary part comes from
the analytic continuation  using $\pm i \epsilon$ and the 
contribution to the spectral
function become  $\delta$-functions at the solutions
of  the dispersion relation. Below the light cone, i.e. when $E<\abs{\vp}$,
there is a finite imaginary part emerging from the logarithms in \eqs{s}{r}
giving a continuous contribution to $\cA(E,\vp)$.
In the appendix it is shown that $S(E,\vp)$ fulfils the general requirements
of normalization and causality.
%
\Subsection{ss:spfknB}{Spectral function in the presence of a $B$-field}
Since the self-energy in the presence of the $B$-field does not
have any singular points away from the real axis, and since
the HTL corrections are negligible for large complex $E$, we expect
that the propagator still has the correct analyticity properties and
that the normalization and causality properties,
discussed in appendix~\ref{a:normcaus}, are
satisfied. We have checked the sum rule in \eq{spnorm} 
by direct numerical calculations
and it is indeed satisfied.
The analytic continuation $E\goto E\pm i\epsilon$~ is more complicated
in the presence of the background field, but it can be summarized by the
formula:
\bea{analcont}
    &&\inv{\sqrt{E^2-p^2}}\arctan\frac{s\sqrt{2qB}}{2\sqrt{E^2-p^2}}\goto
    \nn &&
    \theta(E^2-p^2)\inv{\sqrt{E^2-p^2}}
    \arctan\frac{s\sqrt{2qB}}{2\sqrt{E^2-p^2}}
    \nn && +\,
    \theta(p^2-E^2)\Biggl\{\theta(p_z^2-E^2)\inv{2\sqrt{p^2-E^2}}
    \ln\abs{\frac{2\sqrt{p^2-E^2}-s\sqrt{2qB}}
      {2\sqrt{p^2-E^2}+s\sqrt{2qB}}}
    \nn && \quad\quad\quad\quad +\, 
    \theta(E^2-p_z^2)\left[\frac{\sign(s)}{2\sqrt{p^2-E^2}}
      \ln\abs{\frac{\abs{s}\sqrt{2qB}-2\sqrt{p^2-E^2}}
      {\abs{s}\sqrt{2qB}+2\sqrt{p^2-E^2}}}
    \right.\nn &&\left.
    \quad\quad\quad\quad\quad\quad\quad\quad  
    +\frac{i\pi\sign(s)\sign(E)}{2\sqrt{p^2-E^2}}
    \right]\Biggr\}~~,
\eea
for $E\goto E-i\epsilon$.
\bfig[t]
\setlength{\unitlength}{1mm}
\begin{picture}(50,80)(0,0)
\includegraphics{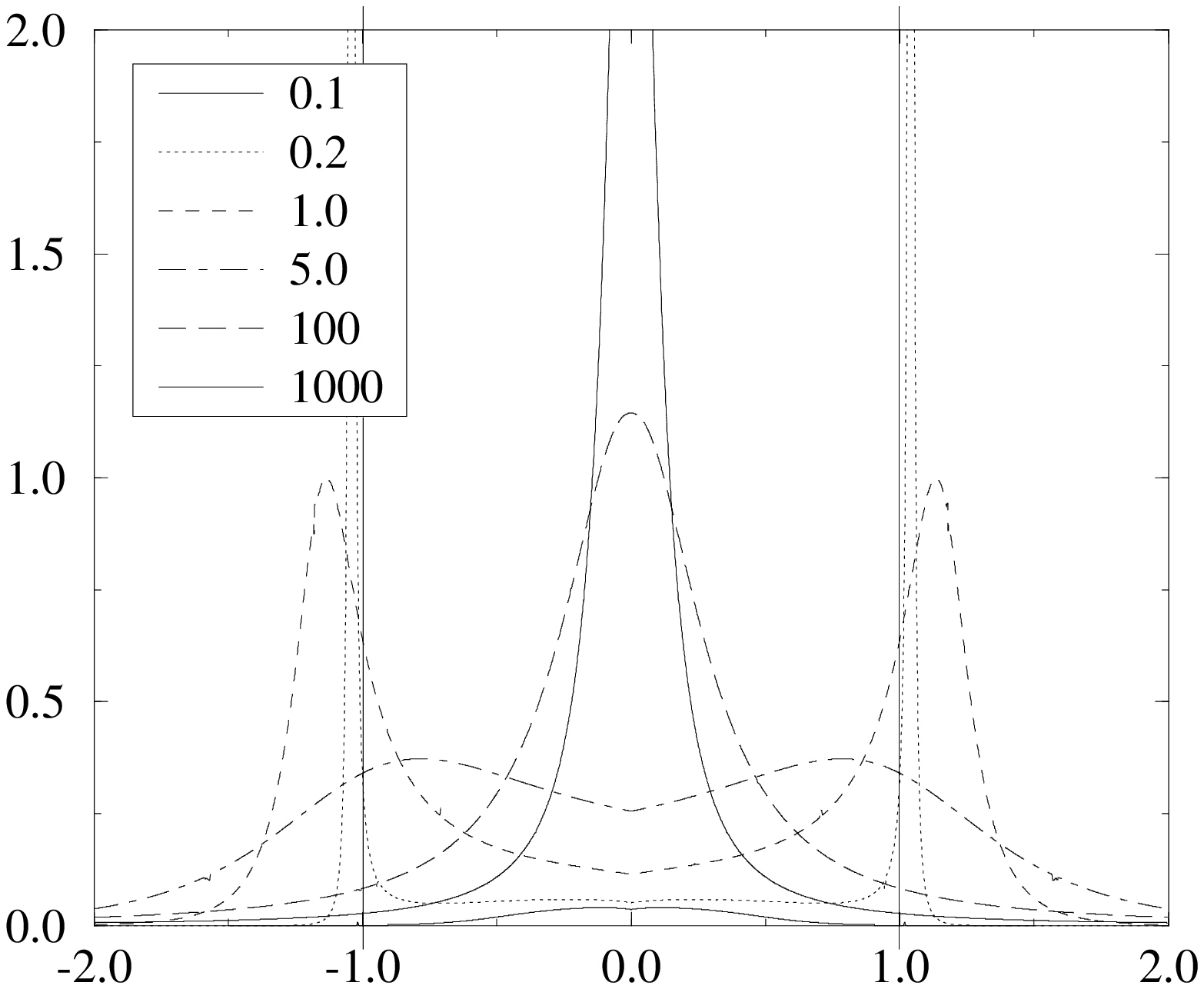}
   \put(0,0){}
   \put(90,-2){$E$}
   \put(17,55){$\cA$}
\end{picture}
   \figcap{Spectral function for various magnetic field strengths
     at the momentum $p_z=0.0$ in the lowest Landau level.
     For weak fields there are two equal peaks around $E\simeq\pm 1.0$
     (indicated by the vertical solid lines),
     corresponding to the thermal mass of the 
     particle and anti-hole solutions. As the
     field increases the width of the peaks increases and the positions
     are shifted to a slightly higher value.
     For intermediate fields the spectral function is very wide
     and eventually it gets more concentrated around $E=0.0$, which 
     is the position it should have without thermal correction.
     All dimensionful parameters are given in units of the thermal
     mass $\cM_e$.
\label{f:sppz0}}
\efig

With this explicit expression it is straightforward to calculate the
spectral function numerically. We have done so for $p_z=0$ and 
$p_z=0.5$ for the right-handed branch 
in the lowest Landau level; the result for
the term proportional to $\g_0$, namely
\bea{ALLL}
    \cA^R_{\rm LLL}(E,p_z)&=&
    \tr\left[\inv 2(1+\g_5)\g_0\cA(E,p_z,n=0)\right]
    \nn
    &=&\inv{2\pi i}\Bigl(S^R_{\rm LLL}(E-i\e,p_z)
    -S^R_{\rm LLL}(E+i\e,p_z)\Bigr)~~,\nn
    S^R_{\rm LLL}(E,p_z)&=&
    \inv{p_0-p_z-\cM_e^2(\<u_0\>-\<u_z\>)}~~,
\eea
is presented 
in \figs{f:sppz0}{f:sppz05}. These figures should be compared
with the solution of the dispersion relation from the real part in
section \ref{s:dr} and correspond to two vertical
cuts in \fig{f:n0cont} at $p_z=0$ and $p_z=0.5$. 
The tendencies are the same. For zero 
momentum (\fig{f:sppz0}) there is no distinction between
particles and holes, and the two $\delta$-function 
peaks correspond to positive and
negative energy solutions. As the
$B$-field is increased, there is a broadening in the width and
the positions of the peaks are shifted towards the zero temperature value,
which at $p_z=0$ is a peak at $E=0$.
At non-zero momentum ($p_z=0.5$, \fig{f:sppz05}) the two peaks at
$B=0$ correspond to a particle solution at $E\simeq 1.1\,\cM_e$ and 
an anti-hole solution at $E\simeq -\cM_e$. 
In addition there is a continuous part in the interval $-p_z<E<p_z$.
Also in this case the peaks get 
broader as $B$ increases and eventually there is only one wide peak around
$p_z=0.5$ for very strong fields.
In vacuum there is an imaginary part of the self-energy, 
describing the decay to a lower
energy level due to synchrotron radiation, only for the 
higher Landau levels, but at finite temperature
even the particle in the lowest Landau level can scatter with the
surrounding plasma and this is what causes the imaginary part.

It should be noticed here that, as already discussed at the end of section
\ref{s:dr}, the HTL approximation is only valid for momenta smaller
than the temperature. For larger momenta, or stronger fields, the 
tree level contribution dominates the real part in the Dirac
equation, but since there is no imaginary part at tree-level
it comes entirely from the HTL term, which is then not reliable
(though small). 
In the very strong field limit ($qB\gg T^2$), 
only the lowest Landau level is occupied
and the imaginary part comes from annihilations with the antiparticles
\cite{ElmforsPS95}.
This term is linear in temperature and does not appear in the HTL 
approximation. 
\bfig[tb]
\setlength{\unitlength}{1mm}
\begin{picture}(50,85)(0,0)
\includegraphics{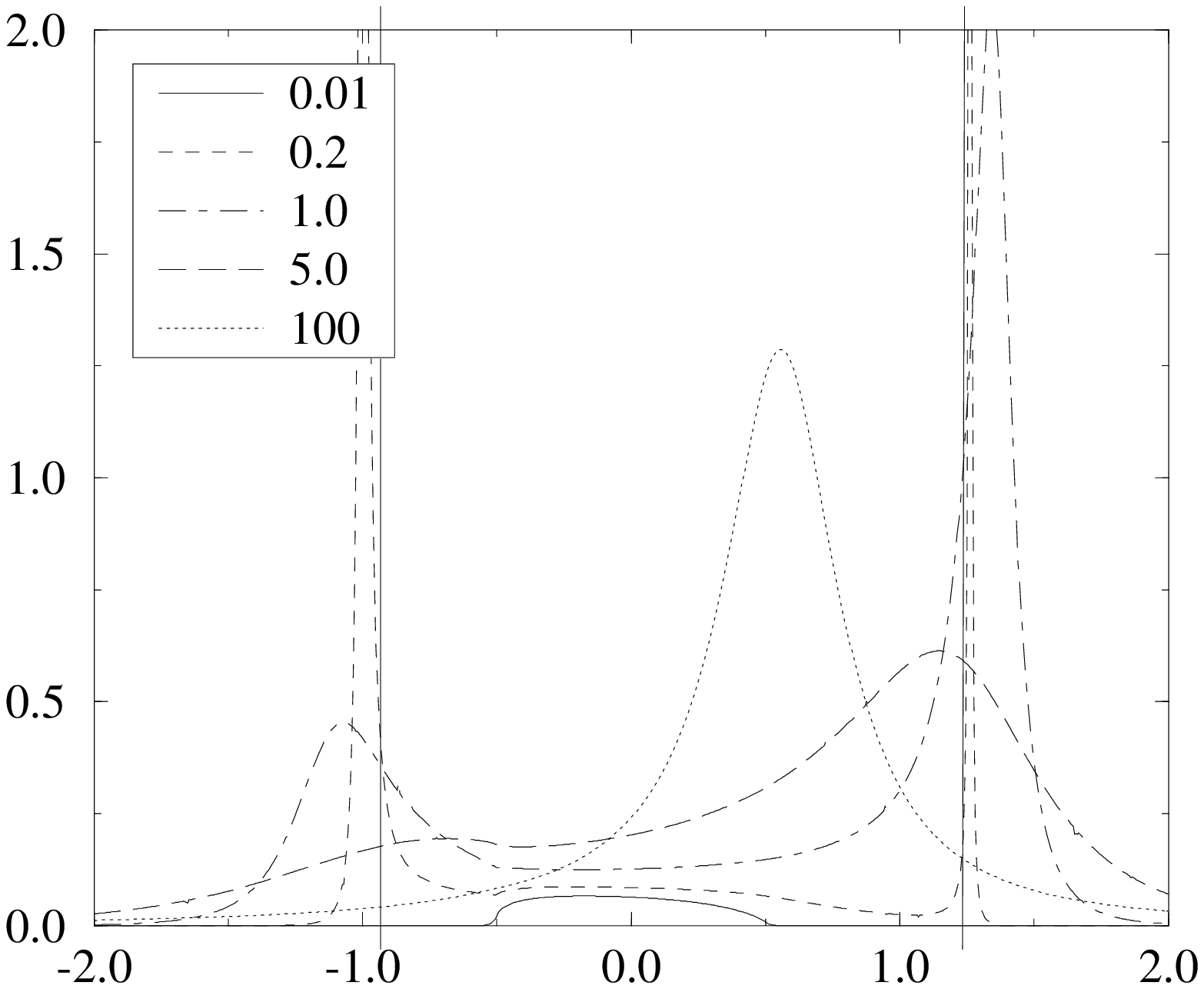}
   \put(0,0){}
   \put(90,-2){$E$}
   \put(17,52){$\cA$}
\end{picture}
   \figcap{Spectral function for various magnetic field strengths
     at the momentum $p_z=0.5$ in the lowest Landau level.
     For very weak fields there are
     two $\delta$-functions at the positions of the particle and 
     anti-hole poles (indicated by the vertical solid lines), 
     and a continuous part in the interval $[-0.5,0.5]$
     which is below the light cone. When the field increases the
     anti-hole peak around $E\simeq -1.0$
     disappears faster than the particle peak at $E\simeq 1.3$.
     At $qB=5.0 \cM_e^2$ there is only a very wide peak left of the
     particle around $E\simeq 1.0$, which for increasing field is
     slowly shifted towards $E=0.5$ while also getting narrower.
\label{f:sppz05}}
\efig
%
\Section{s:chiral}{The chiral anomaly}
The classical 
action for massless fermions is invariant under chiral transformations,
but the corresponding chiral current is not conserved on the 
quantum level due to the chiral anomaly \cite{AdlerBJ69}. 
The divergence of the chiral current in 3+1 dimensions is given by
\be{chcons}
    \pa_\mu j^\mu_5=
    \pa_\mu\Psibar\g^\mu\g_5\Psi=
    \frac{e^2}{16\pi^2}\varepsilon_{\mu\nu\rho\sigma}
    F^{\mu\nu}F^{\rho\sigma}~~.
\ee
Finite temperature effects
do not break chirality and, as a classical action, the HTL
effective action is still chirally invariant. Since the anomalous
term in \eq{chcons} originates from the 
UV-divergent part of the propagator, it is not expected to change
at finite temperature and \eq{chcons} is expected to hold. 
This has also been  verified explicitly by several authors
\cite{ItoyamaM83,Smilga92}.

In vacuum   there is a  clear   physical picture,
related to the IR properties of the fields, of  how  chirality is
created by moving particles  from the Dirac sea  up to positive energy
by     switching     on      an       external     electric      field
\cite{NielsenN83,AmbjornGP83}. This picture works because the massless
dispersion  relation     crosses       the   Dirac      surface    and
particle--antiparticle pairs  can  be  created continuously. By   adding a
chirality-breaking mass term the dispersion relation no longer crosses
the  Dirac surface  and, in   fact, no  chirality  is  produced if the
external gauge field  varies adiabatically. New particle
pairs are again created \cite{AmbjornGP83} when the variation of
the gauge  field is rapid with respect to the  mass of  the fermion. 

At  very high temperature, it is   well known that even massless chiral
fermions pick up  an  effective thermal  mass,  in the sense  that the
energy  of the  propagating  modes  does not   go to zero for  vanishing
momentum. From this IR picture it is not obvious how the anomaly equation
(\eq{chcons}) can be fulfilled at finite temperature. What happens, as
I show below,  is that  as a quasi-particle  of one  chirality moves
along the   dispersion relation from  a particle-like  excitation to a
hole-like one, it looses its spectral weight, while a quasi-particle of
opposite chirality  gains the same  amount of spectral weight. In this
way the spectral weight is shifted between  chiralities (or between states
above and  below the Dirac  surface)  without any dispersion  curve
actually crossing the surface. 

Most discussions of   the anomaly at  zero  temperature  are performed
using a language of single-particle excitations. Even though there are
stable quasi-particles in the leading HTL effective action without any
external field, this  is far  from being the whole  picture. There  is  some
danger in treating the branches of the dispersion relation as ordinary
particles.  Each of the branches does not have a full spectral weight,
and not even  the sum of the spectral  weight for the particle and the
hole adds  up  to 1.  With  non-zero external  field  there are also
imaginary   parts,  which  cannot be    accounted  for within   a
quasi-particle picture. We shall therefore  carry out the calculations
entirely in terms of Green's functions and without reference to 
single-particle states.   The HTL effective action is,
after all, only a way of writing a set of Green's functions. 
%
\Subsection{ss:11anomaly}{The 1+1 D anomaly}
In order to see  what are the essential  parts of the  thermal anomaly
equation, we  shall first briefly repeat  the standard  calculations of
the anomaly in  1+1  dimensions using  both an  operator and a  Green's
functions language.  We shall follow ref.~\cite{AmbjornGP83} very closely in
the operator formalism.  The chiral anomaly at finite temperature for a
non-interacting theory     has  been   studied   by   several  authors
\cite{ItoyamaM83},   in   particular in   1+1 dimensions. 
While  Smilga \cite{Smilga92} gave a  physical
interpretation of  the thermal effect    in terms of  scattering  with
particles in the heat bath, I  take a slightly different approach
to  reach similar conclusions.   Once we have understood the mechanism
in 1+1 dimensions 
it should be easier to see that the same holds true in 3+1 dimensions. 

There is an intrinsic consistency problem in treating a time-dependent
gauge field, i.e.  an electric field,  and an equilibrium ensemble  at
the  same time.  Starting from an   equilibrium ensemble  it  will not
remain  in equilibrium if we  switch on an  electric  field, unless we
consider an  adiabatic limit where the  ensemble has time  to readjust
itself to  equilibrium much faster   than the  variation of the  gauge
field.   
Even though \eq{chcons}  is true at the operator  level, and thus true
for any expectation value  of the equation, it  does not determine the
time-integrated form $\int^t dt'\tr[\rho(t')\Psibar(x)\g_0\g_5\Psi(x)]$
for  an
explicitly  time-dependent density   matrix  $\rho(t)$. The  fields in
\eq{chcons} are  in the Heisenberg picture  and  if the density matrix
has no    explicit  time dependence  it  enters  only  as   an initial
condition.  In  an adiabatic  limit one   could effectively  take into
account interactions with  the heat  bath by  using a  density  matrix,
which at all times corresponds to  thermal equilibrium. This would then
be   an explicitly  time-dependent  density    matrix and the  
time-integrated anomaly equation may not fulfil the standard 
anomaly equation. Such  a
replacement with  an  effective   density  matrix is  to  some  extent
arbitrary and depends  on which physical situation  is imagined. It is
possible to consider  that  only energy is equilibrated  by scattering
processes, and  that chirality is conserved in  each process, and thus
not  equilibrated. Or, one  can consider the  system to  be in contact
with a heat reservoir with which it can also exchange chirality. 

Another possible situation   is  when we neglect  interactions
between particles altogether and follow  the exact time evolution of
the   non-interacting plasma, after  its   initial condition is given.
This is the situation we shall consider in detail.

The chiral charge has to be defined using a gauge-invariant point splitting
regularization in the spatial $z$-direction:
\be{Q5}
    \< Q^\gamma_5\>=
    \int d\xi_z\,d\eta_z\delta_\gamma(\xi_z-\eta_z)
    \<\Psibar(\xi_z,t) \g_0\g_5\Psi(\eta_z,t)\>
    \exp\left[ie\int_{\eta_z}^{\xi_z} A_z(\xi_z',t)d\xi_z'\right]~~,
\ee
where
\be{gdelta}
    \delta_\g(\xi)=\frac{\exp[-\frac{\xi^2}{2\g}]}{\sqrt{2\pi\g}}~~.
\ee
Including a chirality-breaking Dirac mass the anomaly equation
in 1+1 dimensions,
integrated over space and over time from 0 to $\t$, reads
\be{manom}
   \lim_{\g\goto 0}
   \< Q^\gamma_5(\t)\>=
   \frac{e}{2\pi}\int_0^\t \!dt\int\! dx\, \ve_{\mu\nu} F^{\mu\nu}
   +2im\int_0^\t dt\< \overline{Q}^\gamma_5(t)\>~~,
\ee 
where $\< \overline{Q}^\gamma_5\>$ is defined as 
$\<Q^\gamma_5\>$ in \eq{Q5} but with $\g_0\g_5$ replaced by $\g_5$.
With a field operator
\be{Psi}
    \Psi(t,\xi_z)=\int[d{p_z}]e^{i{p_z}\xi_z}
    [u_{p_z}(t)b_{p_z}+v_{-{p_z}}(t)d^\dagger_{-{p_z}}]~~,
\ee
where the notation is taken from \cite{AmbjornGP83}, the chiral charge
in the massless limit can be computed as
\be{Q5Psi}
    \<Q^\gamma_5(\t)\>=
    -\frac{L}{2\pi}\int dp_z \, e^{-\frac{\g}{2}(p_z-eA_z(\t))^2}
    \biggl(1-\<b_{p_z}^\dagger b_{p_z}\>
    -\<d^\dagger_{-{p_z}}d_{-{p_z}}\>\biggr)
    \biggl[\theta({p_z})-\theta(-{p_z})\biggr]~~.
\ee
The initial   expectation      values of the     number  of  particles
$\<b_{p_z}^\dagger          b_{p_z}\>$      and          antiparticles
$\<d^\dagger_{-{p_z}}d_{-{p_z}}\>$ depend on which physical  situation
we consider, but in any case they should go  rapidly to zero for large
$|p_z|$.   In  thermal equilibrium, with  zero  chemical potential, we
would    for      instance             have         $\<b_{p_z}^\dagger
b_{p_z}\>=\<d^\dagger_{-{p_z}}d_{-{p_z}}\>=
(\exp[\beta|{p_z}|]+1)^{-1}$.  It is thus only in the vacuum part that
the  point splitting is needed.  It  follows that only the vacuum part
can  depend  on $A_z(t)$  and  the chirality production is, therefore,
independent of the initial thermal condition.   If, on the other hand,
we  consider a situation where  the particles relax rapidly to thermal
equilibrium, so that the distribution of particles with quantum number
${p_z}$ is  determined by the energy  of the states after switching on
the  $A_z$-field,    then  we  should rather   use  $\<b_{p_z}^\dagger
b_{p_z}\>=(\exp[\beta                E_{p_z}]+1)^{-1}$,          where
$E_{p_z}=|{p_z}|-\sign({p_z})eA_z(t)$. In this  case also  the thermal
part of \eq{Q5Psi} depends    on $A_z(t)$, not directly   through  the
anomaly, but from the interaction with the heat bath that we assume to
be present in order to maintain thermal equilibrium. 

Since in the (3+1)-dimensional case we want to avoid the use of single
particle  states as   in  \eq{Psi}, we shall  now   see how the  above
calculation can   be performed using   Green's functions.   The  field
expectation values can be  related to the time-ordered Feynman Green's
function via 
\be{expS}                              
	\<\Psibar_a(\xi)\Psi_b(\eta)\>=
    -i\at{S_F(\eta,\xi)_{ba}}{\xi_0>\eta_0}~~.  
\ee 
In a  time-dependent background field,  where energy is not conserved,
it  is   not   obvious what  the    correct  $\e$-prescription  for  a
time-ordered  Green's   function     should  be.   In   this    simple
non-interacting  case we can,  however, compute  everything explicitly
and, starting with the vacuum part, we find 
\bea{Gfcn}      \<\bfT\Psi(\xi)\Psibar(\eta)\>^0&=&     \int    [d^2p]
    \exp\left[-ip_0\xi_0+ip_z\xi_z+ i\g_5\int^{\xi_0}dt  A_z(t)\right]
    \nn                     &&                     \frac{i}{p\sslask}~
    \exp\left[ip_0\eta_0-ip_z\eta_z+i\g_5\int^{\eta_0}dt
    A_z(t)\right]~~, 
\eea 
where  the usual Feynman  prescription $p_0\goto (1+i\e)p_0$ should be
used in $\frac{i}{p\sslask}$.    All complications  from  the external
$A_z$  field is  thus  taken  into   account in   the phases  of   the
diagonalizing wave   functions,  and   the  Feynman    prescription is
unchanged. The significance of  $p_0$ depends on the representation of
the wave  functions  we use  to diagonalize  the propagator,  and here
$p_0$ is the initial energy of a particle in a state labelled by $p_z$
before the $A_z$ field  is turned on. The  actual energy of that state
then varies as $p_0-\chi A_z(t)$, where $\chi$ is the chirality of the
state.

With this form  of the  propagator we  obtain  the chiral charge  from
\be{PbggPfin}    
	\<\Psibar(\xi)  \g_0\g_5\Psi(\eta)\>^0=-i\int  [d^2p]
	e^{ip(\xi-\eta)}\at{\left(\frac{e^{-i\int^{\xi_0}_{\eta_0}eA_z(t)dt}}
	{p_0-p_z+i\e p_0}-
	\frac{e^{i\int^{\xi_0}_{\eta_0}eA_z(t)dt}}{p_0+p_z+i\e p_0}
	\right)}{\xi_0>\eta_0}~~.  
\ee 
The  condition $\xi_0>\eta_0$ tells  us
that  the $p_0$ contour  must be closed in  the  upper half-plane. The
poles   in \eq{PbggPfin} give two $\theta$-functions   in $p_z$ in the
standard manner.   The  rest  of the    calculation can be   found  in
\cite{AmbjornGP83} and the result is 
\be{Q5fin} 
	\< Q^\gamma_5(\t)\>^0=
	-\frac{L}{2\pi}\int\!dp_ze^{-\frac{\g}{2}(p_z-eA_z)^2}
	[\theta(p_z)-\theta(-p_z)] \stackrel{\g\goto 0}{\goto} -\frac{L}{2\pi}
	2eA_z(\t)~~, 
\ee 
in accordance with \eq{Q5Psi}.  From this exercise we
learn that in the  massless case, where  we can find an explicit basis
diagonalizing the propagator, the mathematical mechanism that gives us
the correct anomaly is  that the poles  in  $p_0$ cross the real  axes
when $p_z=0$.

In  the  equilibrium real-time finite   temperature formalism the free
propagator can  be written  as in \eq{RTS},  but  since  we compute  a
one-point  function     we only     need  the   11-part:    
\be{Tprop}
	iS_F^{\beta}(p)=iS_F^0(p)
	-f_F(p_0)\left(iS_F^0(p)-iS_F^{0*}(p)\right)~~,
\ee 
where $f_F(p_0)$ is the thermal distribution function.  The problem we
have at hand is not one of an equilibrium, but  as we saw in \eq{Gfcn}
the time dependence can be entirely absorbed in the phases of the wave
functions. Since in this basis $p_0$ has the meaning  of the energy of
the initial state, the thermal version of \eq{PbggPfin} is obtained by
the substitution 
\be{SAsubst}     
	\frac{i}{p_0-\chi    p_z+i\e     p_0}\goto       2\pi
    \sign(p_0)f_F(p_0)\cA^\chi(p_0,p_z)=2\pi
    \sign(p_0)f_F(p_0)\delta(p_0-\chi  p_z)~~, 
\ee 
where $f_F(p_0)$ is
the  initial  particle  distribution.  With   this propagator  the
thermal contribution   to the  anomaly  is given   by  
\be{Q5b} 
	\<Q^\gamma_5(\t)\>^\beta           
	=\frac{L}{2\pi}\int dp_z
    	e^{-\frac{\g}{2}(p_z-eA_z)^2}\int_0^\infty dp_0
    \biggl(f_F(p_0)+f_F(-p_0)\biggr)
    \biggl[\cA^R(p_0,p_z)-\cA^L(p_0,p_z)\biggr]~~,  
\ee  
which  agrees
with the thermal part of \eq{Q5Psi}.  Since the spectral functions 
are rapidly    convergent   in   $p_z$  for  fixed     $p_0$   the
point splitting  $\gamma$  can be sent to  zero   before doing the
integrations.  There is no  $A_z(t)$  dependence left, which shows
again that there is no thermal correction to the anomaly. 
    
\bort{   It is  instructive  to compare  with  the  massive case in an
adiabatically    slowly  varying background  field \cite{AmbjornGP83}.
Since  the   variation of  $A_z(\xi_0)$   is slow  we can   perform  a
derivative   expansion  of    the propagator using     the  method  in
\cite{MossTW92}    \bea{massprop}     \align  S_F(p;\xi)=\int  d^2\eta
e^{ip\eta}S_F(\eta,\xi)                                         \simeq
\frac{p\sslask-eA\sslask+m}{(p-eA)^2-m^2+i\e}           \nn     \align
-ie\frac{p\sslask-eA\sslask+m}{(p-eA)^2-m^2+i\e}               \pa_\mu
A\sslask\left(\frac{\g^\mu}{(p-eA)^2-m^2+i\e}
-\frac{p^\mu(p\sslask-eA\sslask+m)}{(p-eA)^2-m^2+i\e}\right)~~.\nn
\eea We  want to check \eq{manom} to  the zeroth  order in derivatives
but since the right-hand side contains an integral we  need to use the
first order propagator  in $\pa A$   to get the  leading term.   Using
\eq{massprop} in the    left-hand  side of  \eq{manom}  we  find  that
\be{mPbggP} \<\Psibar(\xi,t) \g_0\g_5\Psi(\eta,t)\>=     \int   [dp_z]
e^{-ip_z(\xi^1-\eta^1)}    \frac{p_z-eA_z}{\sqrt{(p_z-eA_z)^2+m^2}}~~.
\ee Now, the  integrand is  shifted in $p_z$   in such a way that  the
total produced chirality is zero. On the right-hand side of \eq{manom}
the anomalous $F\tilde{F}$-terms is, of course, unaffected by the mass
and $\<\overline{Q}_5\>$    gets  only contribution  from    the first
derivative term in \eq{massprop}.  This part is rapidly convergent for
large momenta so $\g$ in the  point splitting $\delta$-function can be
put  to zero.  Finally,    $\<\overline{Q}_5\>$ is exactly   equal but
opposite to  the  anomalous  term,  so the  the total  result  on  the
right-hand side is zero, in agreement with the left-hand side.  } 
%
%
\Subsection{ss:31anomaly}{The anomaly in 3+1 dimensions at high 
temperature} 
The HTL effective action is chirally  invariant even though there is a
mass gap  in the dispersion relation.  We,  therefore, expect that the
chirality produced in  vacuum cannot be undone by chirality-conserving
interaction with the thermal heat  bath. From a mathematical point  of
view we can argue   that since the  anomaly equation  is true at   the
operator level  it  must remain true   in  whatever average  we  take,
including  a thermal  average. We  shall see that  this  is correct by
explicitly  calculating  the chirality   production in  3+1 dimensions
within the HTL approximation.   The only thing we  need is an explicit
expression  for  the propagator.   The  fermionic   part of  the   HTL
effective action is simply related to the inverse of the propagator by
\be{LeqS} 
	\cL_{\rm HTL}^{\rm f}=\Psibar(x) S^{-1}(x,y) \Psi(y)~~.  
\ee
In order   to  write down  the  propagator itself   we have fixed  the
boundary conditions when inverting  the kernel of \eq{LeqS}. At finite
temperature the   standard  inversion   gives   \eq{Tprop},  but  with
\be{S0HTL}                    
	iS_F^0(x,y)=\<\bfT[\Psi(x)\Psibar(y)]\>=
	\left<x\left|\frac{i}{\Pi\slask-m-\cM_e^2                   
	\gamma_\mu
	\left<\frac{u^\mu}{u\cdot\Pi}\right>}\right|y\right>~~.       
\ee  
The $\e$-prescription  for time ordering can  be obtained by comparing the
present  calculation     with   the      explicit   calculation     in
section~\ref{ss:11anomaly}.   We  shall here  use  similar,  explictly
time-dependent, wave functions to diagonalize the propagator. 

The  (3+1)-dimensional anomaly equation at   finite temperature, in  a
classical background field, is given by 
\be{HTLanom}                   
	\<\pa_\mu\Psibar(x)\g^\mu\g_5\Psi(x)\>=
    \frac{e^2}{16\pi^2}\varepsilon^{\mu\nu\rho\sigma}
    F_{\mu\nu}F_{\rho\sigma}~~.  
\ee 
We   shall   compute the left-hand    side   of \eq{HTLanom}  from the
propagator in \eq{S0HTL} in a  background consisting of orthogonal $E$
and $B$ fields.  Scattering with external thermal particles, described
by the thermal  part of \eq{Tprop}, is  discussed at the end.  In  the
massless vacuum case we saw that the essential mechanism of generating
the  correct anomaly was that the  poles of the propagator crossed the
real axis when  $p_z$ changed sign.   This does  not happen  at finite
temperature due to thermal masses. 

First, we need to diagonalize  the HTL Dirac  equation in the presence
of  a $B$-field  in  the $z$-direction and   a parallel  $E$-field. We
choose the gauge  $A_\mu=(0,0,0,A_3(\xi_0))$ for the electric  part of
the background field. The diagonalization can  be done in the same way
as in section~\ref{s:dr} but instead of \eq{vvector} we use the basis 
\be{Evvector}
    \< \xi|v_p\>=\exp\left[-ip_0\xi_0+ie\frac{v_z}{v_0}\int^{\xi_0}A_z(t)dt
      +ip_z\xi_z+ip_y\xi_y+
    i\left(p_x\xi_x+\frac{qB v_y}{v_x}\frac{\xi_x^2}{2}\right)\right]~~,
\ee
but the eigenvalue of $v\cdot\Pi$ remains $v\cdot p$.  In this way the
propagator can be  calculated exactly, even for non-adiabatic  electric
background fields, just  as in the massless  case at zero temperature.
The  prescription for  the  time-ordered  Green's  function is  again
$p_0\goto(1+i\e)p_0$, in this basis.  
We write the    relevant trace  of  the
propagator as 
\be{SFyx}
    \tr S_F(y,x)\g_0\g_5=
    \sum_{\kappa}\sum_{i,j,a=1}^4 \<y,a|\kappa,i\> 
    [S_F(\kappa)\g_0\g_5]_{ij}
    \<\kappa,j|x,a\>~~,
\ee
and use the basis
\bea{Phibasis}
    \align\Phi^{(i)}_a(\xi;p_0,n,p_y,p_z)=\<\xi,a|\kappa,i\>\nn
    \align=e^{-i(p_0\xi_0-p_y\xi_y-p_z\xi_z)}
      \diag [I_{n,p_y}(\xi_x),I_{n-1,p_y}(\xi_x),I_{n,p_y}(\xi_x),
    I_{n-1,p_y}(\xi_x)]_{ab}
    u_b^{(i)}~~,\nn
\eea
where $u_b^{(i)}$ is a set of 4-spinor base vectors, which can be taken
to  be $u_b^{(i)}=\delta_{ib}$.  For $n=0$ there  are only two states,
$u^{(1)}$ and $u^{(3)}$, the others being  identically zero.  With
the  Dirac  operator in  \eq{RLDirac},    diagonalized in the  spatial
quantum numbers, we obtain for the higher Landau levels ($n>0$):
\be{trSkappa}
    \tr S_F(\kappa)\g_0\g_5=\tr D^{-1}_R(\kappa)-\tr D^{-1}_L(\kappa)~~.
\ee
In the lowest Landau level the matrices $D_{R,L}$ are not invertible,
since there are only two states in total. Its explicit form then is 
($\kappa_0=\{n=0,p_y,p_z\}$):
\be{SkappaLLL}
    \tr S_F(\kappa_0)\g_0\g_5=
    \int_{-\infty}^\infty \frac{\cA^R_{\rm LLL}(E,p_z)-
      \cA^L_{\rm LLL}(E,p_z)}{p_0-E+i\e p_0}~~,
\ee
with $\cA^{R,L}_{\rm LLL}(E,p_z)$ given by
\be{RLLLsp}
    \at{\inv{p_0-p_z-\mme(\<u_0\>_{\kappa_0}-\<u_z\>_{\kappa_0})}}
	{p_0\goto p_0+i \e p_0}
    =\int_{-\infty}^\infty dE\frac{\cA^R_{\rm LLL}(E,p_z)}{p_0-E+i\e p_0}~~,
\ee
and $\cA^{L}_{\rm LLL}(E,p_z)=\cA^{R}_{\rm LLL}(E,-p_z)$.
We shall start by computing the contribution to $\<Q_5\>$ from the 
lowest Landau level. This is, in fact, the only part that contributes,
as we shall see later. Taking the point splitting only in the
$z$-direction we find
\be{Q5LLL}
    \<Q_5\>_{\rm LLL}=-iV\int d\xi_z \delta_\g(\xi_z)
    \int [dp_0][dp_y][dp_z] \<0|\kappa_0\>\tr S_F(\kappa_0)\g_0\g_5
    \<\kappa_0|\xi\> e^{ieA_z\xi_z}~~.
\ee
Using
\be{Iint}
    \int[dp_y]I_{n,p_y^2}(\xi_x)=\frac{eB}{2\pi}~~,
\ee
and the fact that the $p_0$-contour should be closed in the upper half-plane,
the produced chiral charge reduces to
\be{Q5Z}
    \<Q_5\>_{\rm LLL}=-\frac{VeB}{4\pi^2}\int dp_z 
    e^{-\frac{\g}{2}(p_z-eA_z)^2}
    \left[W(p_z)-W(-p_z)\right]~~.
\ee
This equation has a clear resemblance with \eq{Q5fin}.
The function $W(p_z)$ is defined by
\be{Wdef}
    W(p_z)=\int_0^\infty dE \cA^R_{\rm LLL}(E,p_z)~~.
\ee
It is the spectral weight for the right-handed positive energy solution
in the lowest
Landau level. For very large $|p_z|$ there are no collective
excitations, such as holes, but only the standard particle solution. 
With our convention that the $B$-field points in the positive
$z$-direction, we find that for $p_z>0$, $\cA^R_{\rm LLL}(E,p_z)$
is concentrated at a positive energy particle solution at $E=|p_z|$,
while for  $p_z<0$ it is peaked on a negative energy antiparticle
solution at $E=-|p_z|$ (see \fig{f:spcont}). 
\bfig[tb]
\setlength{\unitlength}{1mm}
\begin{picture}(50,100)(0,0)
\includegraphics{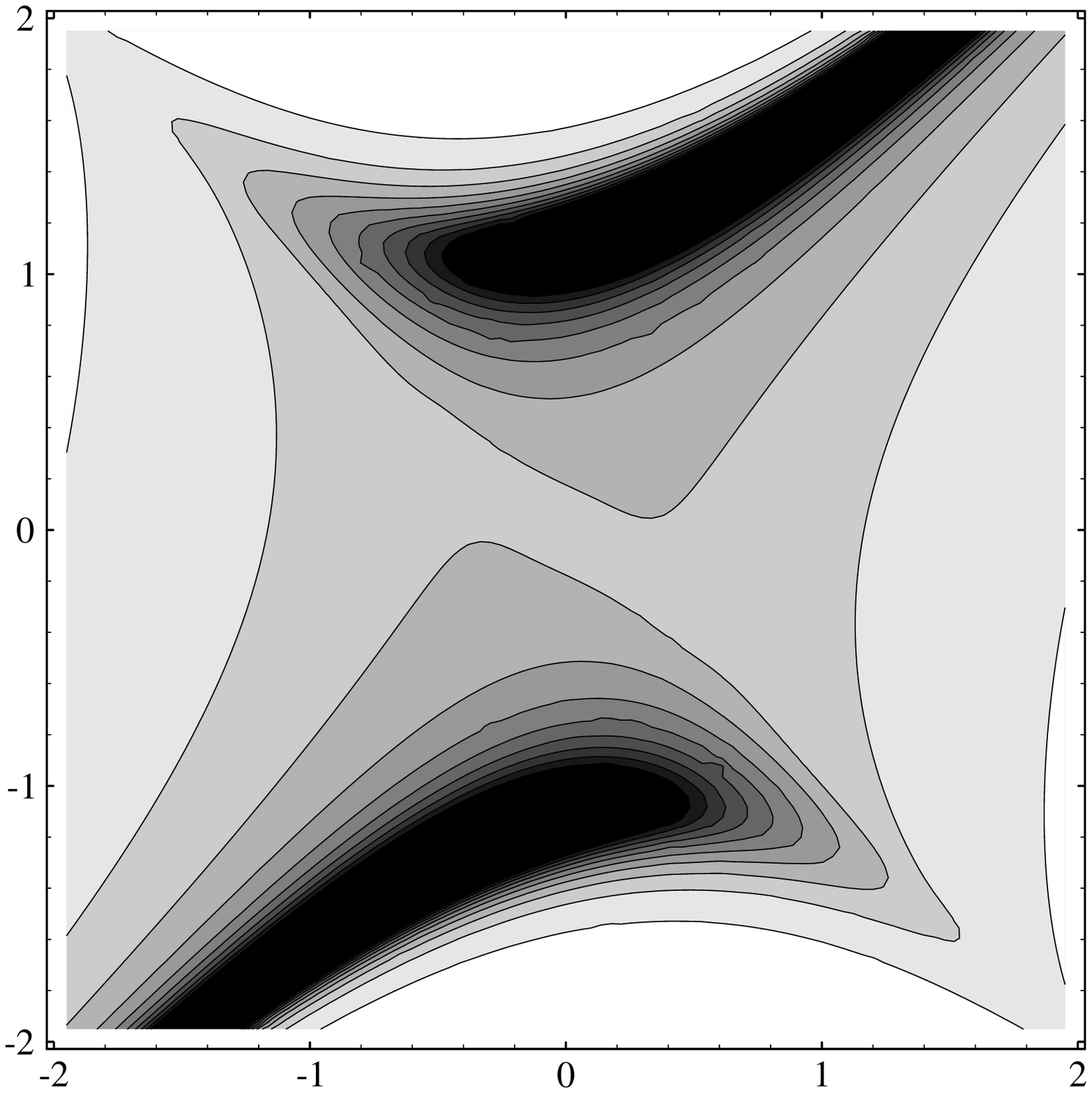}
   \put(0,0){}
   \put(93,-2){$p_z$}
   \put(17,65){$p_0$}
\end{picture}
   \figcap{A density plot of the spectral function 
     $\cA^R_{\rm LLL}(p_0,p_z)$ for $qB=\cM_e^2$.
     Even though there is no crossing of the Dirac surface, the spectral
     density goes continuously between positive and negative energy
     states when $p_z$ decreases.
\label{f:spcont}}
\efig
Thus, we have
\be{limW}
   \lim_{p_z\goto-\infty}W(p_z)=0~,\quad\lim_{p_z\goto\infty}W(p_z)=1~~,
\ee
and all derivatives of $W(p_z)$ vanish for large $|p_z|$.
It can then be shown that 
\be{D5fin}
    \<Q_5(t)\>_{\rm LLL}=-\frac{VeB}{2\pi^2}eA_z
    =\int d^3x\int^tdt'\frac{e^2}{16\pi^2}
    \varepsilon^{\mu\nu\rho\sigma}F_{\mu\nu}F_{\rho\sigma}~~.
\ee
We  note that this   agrees with the anomaly   in \eq{HTLanom} for the
particular  background field configuration  that we have chosen.  When
it comes to  the higher Landau levels it  turns out that the two terms  in
\eq{trSkappa}  are  separately  well convergent   for  large  $p_z$, as
opposed to \eq{SkappaLLL}, which needs the  point splitting in order to
be  well defined. We  can,    therefore, change  $p_z\goto   -p_z$  in
$D_L(\kappa)$, after which the sum cancels when $\g\goto 0$. 

Scattering with particles in the thermal heat bath is taken into 
account in the same way as in 1+1 dimensions. The thermal part is 
UV-convergent and, exactly as in \eq{Q5b}, it has no time dependence.
%
\Section{s:concl}{Conclusions}
The main computational part of  this  paper is the diagonalization  of
the  fermionic part of the Hard  Thermal Loop effective  action in the
presence  of a constant   background magnetic   field.  This  makes  it
possible to  write down the explicit  expression for the spectral function
of  fermions and to  see how it depends on the
magnetic field strength.  We find that, starting from weak fields, the
spectral weight moves from the standard particle and hole solutions at
high temperature   over to the  vacuum Landau  levels  for very strong
fields.  

It has  been recognized  in the  literature  that it  is difficult  to
reconcile the standard  picture of level  crossing as a  mechanism for
anomalies, with the  thermal masses of   fermions at high  temperature
\cite{Smilga92}.  Using  the exact spectral  function, I have shown in
this paper  how   the spectral weight   can  move continuously between
chiralities, when a background electric field  is switched on, without
the  dispersion  relation   ever   crossing the   Dirac  surface.  The
difference from the vacuum is that the spectral weight on a dispersion
curve varies   continuously   between  zero   and   one  in  the   HTL
approximation,  while    it is always exactly  one    in  vacuum.  The
conclusion  is   that the  anomaly   equation  remains valid  at  high
temperature,  even  after taking   interactions into  account.  It  is
however possible to obtain  different production rates by coupling the
system to an external heat reservoir, which means effectively using an
explicitly time-dependent density matrix.  The chirality production in
that case would  then depend on   the exact experimental setup,  and I
have not discussed this possibility in any detail. 

Even though the main problems formulated  in this paper have also been
solved  here,   there are some  related  issues  that still   call for
solutions.  One problem is  to extend  the  analysis of the dispersion
relation to non-Abelian gauge  bosons, but this  turns  out to be  far
more  complicated due  to    self-interaction.  Another problem   of a
certain  interest is to see how  this anomaly mechanism  fits into the
language  of index   theorems, which  has shown  to  be  useful for the
anomalies at zero temperature.

\section*{Acknowledgements}
I would like to thank Chris Korthals-Altes for 
numerous detailed discussions.
\appendix
\Section{a:normcaus}{Normalization and causality}
The thermal expectation value of the
canonical anticommutation relation for the fermionic fields
\be{car}
    \<\{\Psi(x),\Psi^\dagger(y)\}\>\equiv C(x-y)
\ee
should vanish for space-like $x-y$ and should be equal to a $\delta$-function
in $\vx-\vy$ when $x_0=y_0$. These are basic requirements of the fundamental
fields which we do not expect to be violated by the HTL approximation.
In terms of the spectral function the normalization condition, derived from
the equal-time commutator, becomes
\be{spsumrule}
    \int_{-\infty}^\infty dE \cA(E,\vp)=
    \inv{2\pi i}\int_{-\infty}^\infty dE 
    \Bigl(S(E-i\eps,\vp)-S(E+i\eps,\vp)\Bigr)=\g_0~~.
\ee
There are two ways of showing the validity of \eq{spsumrule} for the 
propagator in \eq{SHTL}, analytically and numerically. 
First we present the analytic proof.
Consider the contribution from
the advanced and the retarded propagators separately. 
For a given $\vp$ they are analytic
in the lower and upper half-plane, respectively \cite{BrosB96}, 
and the integration contour can be 
deformed to suitable arcs at infinity. The integrals along those arcs do not
vanish, but the HTL corrections go to zero. The expression for a free
fermion can then be used and it trivially satisfies \eq{spsumrule}. 
In this way it is clear that the only important 
ingredients for the spectral sum rule to be fulfilled is the analyticity
 away from the real $E$-axis, and that the correction goes away for large
complex $E$. It is only the $\g_0$-part of the propagator in \eq{SHTL}
that contributes
to \eq{spsumrule} since the other parts of \eq{SHTL} decay too fast
on the arcs at infinity, and they are also antisymmetric in $E$.
It is, therefore, common to write the normalization condition as
\be{spnorm}
    \int_{-\infty}^\infty dE\inv 4\tr[\gamma_0\cA(E,\vp)]=1~~.
\ee

The other way to check \eq{spnorm} is by a direct
numerical calculation. The residues at the poles above the
light cone ($E>|p|$) have been computed by several
authors \cite{Weldon8289,Petitgirard91}
and it is well known that they do not add up to 1
for $\abs{\vp}>0$. It is straightforward to calculate the integral
in \eq{spnorm}  below the light cone, and it turns out to make up
for the missing part, as expected.

The causality condition, \ie that $C_F(x)$ vanishes for space-like $x$,
has been discussed in the HTL approximation in \cite{HenningPSB95}, where 
\be{CF}
    C(x)=
    \int\frac{dEd^3\vp}{(2\pi)^3}
    \exp[-i(Et-\vp\vx)]\cA(E,\vp)
\ee
was calculated numerically for gauge bosons.
I shall here give an analytic demonstration that the commutator
indeed vanishes outside the light cone,
starting with the part of \eq{CF} proportional to $\g_0$,
$C_0(x)\equiv \tr\inv 4 \g_0 C(x)$.
Assuming that $\tr[\g_0\cA(E,\vp)]$ only depends on $p=|\vp|$, we can perform
the angular integral and obtain
\bea{CFang}
    C_0(x)&=&
    \inv{(2\pi)^2}\int_{-\infty}^\infty dE\int_0^\infty dp\,p^2
    e^{-iEt}\frac{e^{ip|x|}-e^{-ip|x|}}
    {ip|\vx|}\tr[\inv 4\g_0 \cA(E,p)]
    \nn
    &=& -\inv{4\pi^2|\vx|}\frac{d}{d|\vx|}\int_{-\infty}^\infty dE\,dp\,
    e^{-iEt+ip|x|}\tr[\inv 4\g_0 \cA(E,p)]~~.
\eea
In order to more easily see the analytic structure, we change the variables to
$u$ ($=E+p$) and $v$ ($=E-p$):
\bea{CFuv}
    C_0(x)&=&-\inv{8\pi^2|\vx|}\frac{d}{d|\vx|}
    \int_{-\infty}^\infty du\,dv
    \exp\left[-\frac{i}{2}u(t-|\vx|)-\frac{i}{2}v(t+|\vx|)\right]
    \nn&&\times
    \inv{2\pi i}\tr\left[\inv 4\g_0\Bigl(S(u-i\epsilon,v-i\epsilon)-
      S(u+i\epsilon,v+i\epsilon)\Bigr)\right]~~.
\eea
Let us first see how it works for a free massless scalar, where
\be{S0}
    S(E-i\epsilon,p)=\inv{E^2-p^2-iE\epsilon}=
    \inv{u-\frac{i}{2}\epsilon}\,\inv{v-\frac{i}{2}\epsilon}~~.
\ee
The integrals over $u$ and $v$ factorize and we can use 
\be{theta}
  \int_{-\infty}^\infty
  du\frac{e^{i\alpha u}}{u-i\epsilon}=2\pi i\theta(\alpha)~~,
\ee
to show that%
\footnote{Note that there is a sign error in the revised version of 
ref.~\cite{HenningPSB95}.}
\bea{C0}
   C_0(x)&=&-\frac{i}{4\pi|\vx|}\frac{d}{d|\vx|}
   \left[\theta(-t+|\vx|)\theta(-t-|\vx|)-
     \theta(t-|\vx|)\theta(t+|\vx|)\right] \nn
   &=&-\frac{i}{4\pi|\vx|}
   [\delta(t-|\vx|)-\delta(t+|\vx|)]
   =-\frac{i}{2\pi}\sign(t)\delta(t^2-|\vx|^2)~~.
\eea
From this follows also the canonical commutation relation for scalar
fields:
\be{CCR}
    [\phi(t,\vx),\pa_t\phi(t,\vy)]=\pa_{y_0} \at{C_0(x-y)}{y_0\goto x_0}
    =i\delta^{(3)}(\vx-\vy)~~.
\ee
The basic reason why the commutator vanishes outside the light cone is the
occurrence of $\theta$-functions coming from \eq{theta}
and  this follows from
the property of analyticity in the correct 
half-plane. Since $S_0(u-i\epsilon,v-i\epsilon)$ is analytic in the 
lower half-plane for both $u$ and $v$, and it vanishes fast enough for large
arguments, the integration contours can be closed in the lower half-plane
for positive $t-|\vx|$ or $t+|\vx|$, respectively. We conclude that for any 
$S(u-i\epsilon,v-i\epsilon)$ with the correct analyticity properties
we have, using $\theta(\alpha)\theta(\beta)=\theta(\alpha\beta)\theta(\beta)$,
\be{Rtheta}
    \int_{-\infty}^\infty\!du\,dv 
    \exp\left[-\frac{i}{2}u(t-|\vx|)-\frac{i}{2}v(t+|\vx|)\right]
    S(u-i\epsilon,v-i\epsilon)
    = \theta(t^2-|\vx|^2)F(t,|\vx|)~~,
\ee
for some function $F(t,|\vx|)$,
at least for non-zero $t^2-|\vx|^2$.
There can be other singularities right on the light cone.
A similar argument applies to
$S_0(u+i\epsilon,v+i\epsilon)$, leading to another factor
$\theta(t^2-|\vx|^2)$.

Let us return to the HTL propagator and see if it has the correct
analyticity properties. It is given by \eq{SHTL}, and 
in terms of $u$ and $v$ we have, for $E\goto E-i\epsilon$:
\bea{sandruv}
    s(u-i\e,v-i\e)&=&1-\frac{2\cM_e^2}{u^2-v^2}
    \ln\left(\frac{u-i\e}{v-i\e}\right)~~,\nn
    r(u-i\e,v-i\e)&=&1+\frac{4\cM_e^2}{(u-v)^2}
    \left(1-\frac{u+v}{2(u-v)}
      \ln\left(\frac{u-i\e}{v-i\e}\right)\right)~~.
\eea
The cuts from the logarithm start slightly above the real axis and the
branch  cuts remain  in the upper   half-plane.  There is a  potential
singularity  at  $u=v$, but it   can be shown  that both  $s(u,v)$ and
$r(u,v)$ have power series expansion  around that point. Thus, none of
these functions have any non-analyticity in the lower half-plane. Then
we only have to verify that the denominator in \eq{SHTL} does not have
any pole in the lower half-plane. This should be done  for each of $u$
and $v$ keeping the other one real, and it is not very difficult to do
this numerically.  With a  fine enough grid  one can demonstrate  that
there are  no singularities  away  from the real axis.   When deriving
\eq{CFang}, we  assumed that the  spectral  function only  depended on
$|\vp|$.  The  HTL spectral function also  has  a term proportional to
$\vg\vp$, but  it can be  rewritten  as $i\vg\cdot\nabla$  acting on a
rotationally invariant function, so that it does not really affect the
above reasoning. 

\end{document}